\documentclass[letterpaper,twocolumn,10pt]{article}
\usepackage{dsfont}
\usepackage{tikz}
\usepackage{graphicx}
\usepackage{ifpdf}
\usepackage{latexsym}
\usepackage{paralist}
\usepackage{comment}
\usepackage{xspace}
\usepackage{mathrsfs}
\usepackage{setspace}
\usepackage{algorithm}
\usepackage[noend]{algorithmic}
\usepackage{amsmath}
\usepackage{amsfonts}
\usepackage{amssymb}
\usepackage{textcomp}
\usepackage{stfloats}
\usepackage{verbatim}
\usepackage{array}
\usepackage{leftidx}
\usepackage{cases}
\usepackage{multirow}
\usepackage{bbding}
\usepackage{booktabs}
\usepackage{float}
\usepackage{makecell}
\usepackage{threeparttable}
\usepackage{tabularx}
\usepackage{usenix-2020-09}
\newcommand{\paraspace}{\vspace{0.05in}}
\newcommand{\parab}[1]{\paraspace\noindent{\bf #1} }

\begin{document}

\def\mytitle{\textit{TopoEP}: Topology-Aware Load Balancing for Expert-Parallel MoE Training}
\newcommand{\submissiontrack}{Traditional Research Track}

\date{}
\title{\Large \bf \mytitle}
\author{
{\rm Jiacheng Zhu, Xie Zhao, Gongming Zhao,}\\
{\rm Hongli Xu, Yao Fei, and Jin Fang}\\
University of Science and Technology of China\\
{\small\normalfont\ttfamily
\{zhu\_jc,zhaoxie,yao\_fei,fangjin98\}@mail.ustc.edu.cn}\\
{\small\normalfont\ttfamily
\{gmzhao,xuhongli\}@ustc.edu.cn}
}

\maketitle


\begin{abstract}
Dynamic routing creates severe load imbalance in large-scale expert-parallel Mixture-of-Experts (MoE) training, turning GPUs that host hot experts into stragglers.
As each MoE layer waits for its slowest rank, these stragglers prolong the expert-parallel stage and reduce overall training efficiency.
Existing expert-parallelism load-balancing (EPLB) systems commonly compute load-balancing plans on the CPU, incurring device--host data transfers and cross-rank synchronization that make scheduling at every layer and microbatch expensive.
Their planning formulations also overlook the hierarchical communication costs of modern scale-up and scale-out GPU clusters.

We present \textit{TopoEP}, a GPU-native, topology-aware load-balancing system for large-scale MoE training.
At each MoE layer and training microbatch, \textit{TopoEP} converts the current routing result into hot-expert replication and token-rerouting decisions and executes the resulting plan without data-dependent host synchronization, reducing critical-path overhead.
To generate these decisions, \textit{TopoEP} uses a deterministic GPU solver that performs inter-node placement followed by intra-node refinement, allowing all ranks to independently produce bitwise-identical plans.
On a 32-GPU NVIDIA H800 cluster, integrating \textit{TopoEP} with Megatron-LM improves end-to-end training throughput by 6.2\%--11.4\% across three representative MoE models.
\end{abstract}

\section{Introduction}\label{sec:intro}
Mixture-of-Experts (MoE) is a widely adopted architecture for scaling large
language models (LLMs). By activating only a small subset of experts for each
token, MoE increases model capacity without proportionally increasing
per-token computation~\cite{shazeer2017outrageously,lepikhin2021gshard,
fedus2022switch,dai2024deepseekmoe,deepseekai2024deepseekv3}. Expert
parallelism (EP) enables large-scale MoE training by partitioning experts
across GPUs. The \textit{dispatch} phase sends token activations to the ranks
hosting the router-selected experts, and the \textit{combine} phase returns
their outputs to the source ranks~\cite{lepikhin2021gshard,gale2023megablocks,
rajbhandari2022deepspeedmoe,hwang2023tutel}. EP has therefore become a standard
strategy for distributing expert parameters and computation across ranks in
large-scale MoE systems.

As EP groups grow, rank-level load imbalance can become a major system
bottleneck~\cite{pmlr-v235-kim24w}. Top-$k$ gating may concentrate token--expert assignments on a small
set of hot experts whose identities and loads vary across layers and
microbatches~\cite{wei2026ultraep,skiadopoulos2026symi,nie2022evomoeevolutionalmixtureofexpertstraining}. Under static expert placement, this expert-level skew can
translate into uneven aggregate loads across ranks, making ranks that host hot
experts stragglers. Because an MoE layer completes only after every
participating rank finishes, even one overloaded rank can prolong the entire
EP stage. The same skew can also produce imbalanced
\textit{dispatch}\,/\,\textit{combine} traffic, creating communication
hotspots and further reducing end-to-end training
efficiency~\cite{wang2024auxiliary,nie2023flexmoe,deepep2025}.

Existing approaches address this imbalance at either the model or system level.
Model-level approaches modify routing through auxiliary losses, routing biases or capacity-induced token dropping, potentially affecting model quality~\cite{lepikhin2021gshard,fedus2022switch,wang2024auxiliary}.
System-level methods preserve each token's logical expert assignments while replicating hot experts and redistributing their assigned tokens across expert instances~\cite{he2022fastermoe,deepseekai2025eplb,nguyen2026leastloaded}.
Such methods need to generate each plan rapidly because planning lies on the critical path before token dispatch.
Existing online approaches commonly rely on host-side optimization solvers, whose GPU--CPU synchronization, CPU solving and plan distribution add directly to this critical-path latency.
More recent systems such as UltraEP~\cite{wei2026ultraep} and
MoonEP~\cite{moonep2026} adopt GPU-native load balancing but target a single
scale-up domain, overlooking the heterogeneous costs of intra-node NVLink and
inter-node RDMA. Consequently, when extended to scale-out, topology-unaware
planning may reduce rank-load imbalance while incurring costly cross-domain
token, expert-parameter, and replica-gradient transfers, offsetting the
performance gains from improved load balance.
These limitations motivate an online system-level load balancer that improves training efficiency without
altering Top-$k$ expert selection or compromising model quality.

Realizing such a system presents three coupled challenges: planning quality, solver latency and execution overhead.
First, the planner needs to derive expert-replication and token-rerouting decisions from the current Top-$k$ gating result under a limited replica budget.
A high-quality plan reduces the maximum expert-computation time across ranks while accounting for communication costs.
Second, plan generation lies on the critical path between Top-$k$ gating and token dispatch, leaving little opportunity to hide its latency.
At large EP scales, rapidly solving this joint replication-and-routing problem across many experts and ranks is challenging because any solver delay directly postpones token dispatch and can offset the gains from improved load balance.
Finally, applying each plan introduces expert-parameter transfers, replica-gradient aggregation and additional memory usage.
These overheads need to remain low enough for improved load balance to translate into end-to-end performance gains.

We therefore present \textit{TopoEP}, a GPU-native load-balancing system that transforms the routing outcome of each MoE layer and microbatch into an expert-replication and token-rerouting plan.
For high-quality planning, \textit{TopoEP} jointly determines
replica placement and token allocation among expert instances,
subject to a per-rank replica limit.
Inter-node placement reduces RDMA traffic, while intra-node
refinement mitigates residual rank-load imbalance.
For low-latency solving, \textit{TopoEP} exploits block-level concurrency, shared-memory caching and warp-level reductions to execute this deterministic procedure entirely on the GPUs.
Given the same global routing matrix and configuration, every rank independently produces a bitwise-identical plan without data-dependent host coordination or an additional plan broadcast.
For low-overhead execution, \textit{TopoEP} uses reusable replica buffers and device-initiated communication through TMA~\cite{nvidia_hopper_2022} and NCCL GIN~\cite{hamidouche2025gpu}.
It orchestrates the compute and communication streams through a two-chunk pipeline to reduce exposed communication overhead.

We integrate \textit{TopoEP} into the Megatron-LM training framework and
evaluate it on a 32-GPU NVIDIA H800 cluster. Compared with the Megatron-LM
baseline, \textit{TopoEP} achieves lower rank-load imbalance,
reduces cross-domain token--expert assignments by 96.2\%--98.8\%, and improves
end-to-end training throughput by 6.2\%--11.4\% across three representative
MoE models, without observable degradation in training-loss convergence.

In summary, this paper makes the following contributions:

\begin{itemize}
    \item \textbf{Topology-aware deterministic GPU planning.}
    We formulate joint replica-placement and token-routing decisions across
    scale-up and scale-out fabrics and develop a two-stage GPU algorithm that
    trades cross-domain token traffic against replica-transfer cost before
    refining rank loads within each domain. Its deterministic parallel
    execution lets every rank independently generate an identical plan.

    \item \textbf{GPU-resident plan execution.}
    \textit{TopoEP} applies each plan using reusable replica buffers,
    device-initiated TMA\,/\,GIN communication, and a two-chunk pipeline,
    reducing the critical-path overhead of parameter and gradient transfers as
    well as token communication.

    \item \textbf{End-to-end system and evaluation.}
    We integrate \textit{TopoEP} into Megatron-LM and evaluate solver
    scalability, load balance, communication locality, end-to-end throughput,
    and training stability on a cluster of 32 NVIDIA H800 GPUs.
\end{itemize}

\section{Background and Motivation}\label{sec:background}
\subsection{Expert-Parallel MoE Execution}\label{subsec:ep-execution}

A dense Transformer layer applies the same feed-forward network (FFN) to every token~\cite{vaswani2017attention}. In contrast, an MoE layer contains multiple independently parameterized FFNs, called experts. For each token, a trainable router selects the Top-$k$ experts based on its hidden representation. By activating only a small subset
of experts, MoE increases model capacity without a proportional increase in
computation~\cite{shazeer2017outrageously,lepikhin2021gshard,fedus2022switch}.

As the number and size of experts increase, a single GPU can no longer store and execute every expert in an MoE layer. Large-scale MoE models therefore use expert parallelism (EP)~\cite{lepikhin2021gshard}, which partitions experts across multiple GPUs. A token
activation originating on one GPU may consequently be routed to an expert hosted on another, distributing both expert parameters and computation across
devices~\cite{lepikhin2021gshard,rajbhandari2022deepspeedmoe}.

\begin{figure}[!htbp]
    \centering
    \includegraphics[width=\columnwidth]{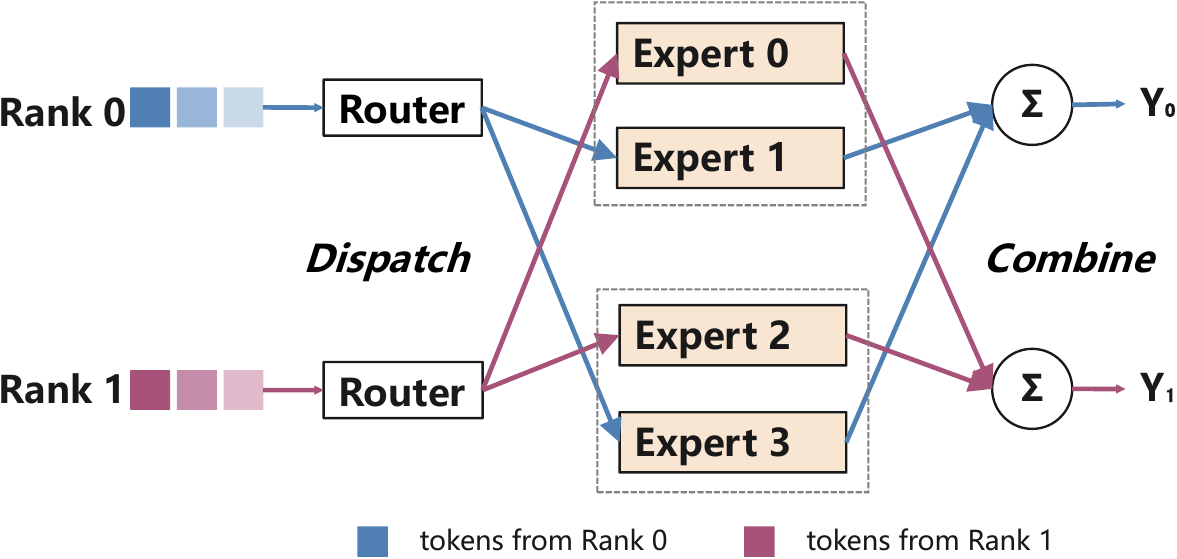}
    \caption{Execution flow of an expert-parallel MoE layer.}
    \label{fig:ep-execution}
\end{figure}

As shown in Figure~\ref{fig:ep-execution}, each rank first applies the router to
its local tokens. An all-to-all dispatch sends the resulting token activations
to the ranks hosting the selected experts, where the experts evaluate their
assigned tokens~\cite{hwang2023tutel}. Since expert computation is dominated by
matrix multiplications, its cost generally grows with the number of assigned
tokens~\cite{ICLR2026_6ed5bf44,MLSYS2025_e27ea0cd}. A reverse all-to-all then returns the expert outputs to their source
ranks, where they are combined using the corresponding routing weights.

The realized expert load becomes available only after routing. An online planner that uses this load must therefore make the resulting plan available to all participating ranks before dispatch begins. Any exposed planning latency directly delays dispatch
and expert computation, placing the planner on the critical path. This timing
constraint is central to
Section~\ref{subsec:online-eplb-critical-path}.

\subsection{Dynamic Expert Load Imbalance}
\label{subsec:dynamic-load-imbalance}

Routing decisions depend on each token's hidden representation, so expert loads
can be skewed and vary with the input~\cite{pmlr-v162-liu22g,zhou2022mixtureofexpertsexpertchoicerouting}. When a batch contains many tokens
associated with a small set of learned patterns, the corresponding experts may
be selected disproportionately often. These hot experts receive substantially
more assignments than the mean, while other experts receive few or none.
We profile all 48 MoE layers of Qwen3-30B-A3B~\cite{qwen3technicalreport} on two datasets: DAPO-Math~\cite{yu2025dapo} for mathematical reasoning and StarCoderData~\cite{li2023starcoder} for code. Each layer contains 128 logical experts and routes each token to eight experts.

\begin{figure}[!htbp]
    \centering
    \includegraphics[width=\columnwidth]{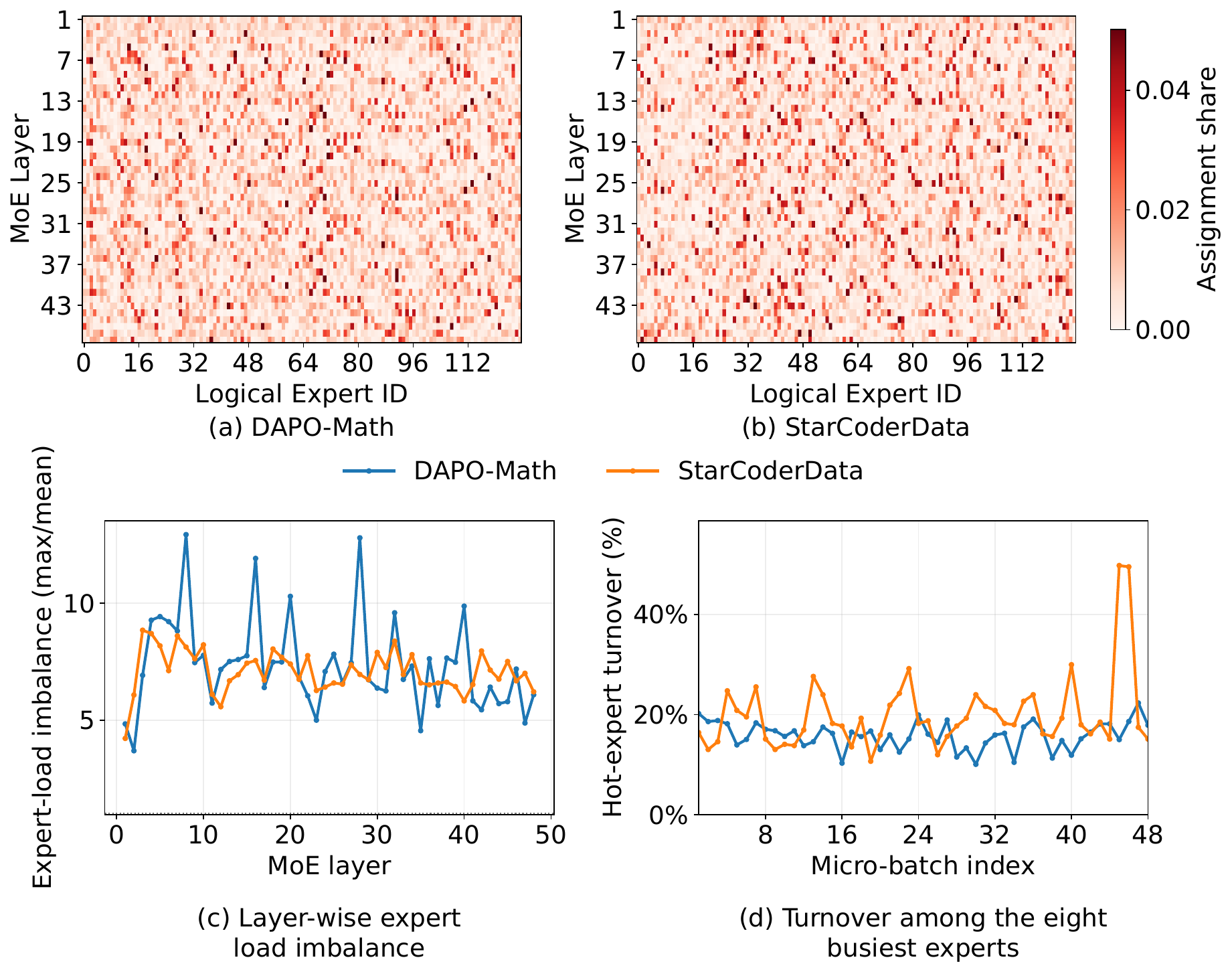}
    \caption{Expert-load skew across the 48 MoE layers of Qwen3-30B-A3B on
    DAPO-Math and StarCoderData. (a)--(b) Fraction of each layer's token--expert
    assignments routed to each expert (horizontal axis: logical expert ID
    0--127; vertical axis: MoE layer 1--48). (c) Per-layer maximum-to-mean
    expert-load ratio. (d) Fraction of the eight highest-load experts that
    changes between adjacent microbatches.}
    \label{fig:expert-load-hotspots}
\end{figure}

Figure~\ref{fig:expert-load-hotspots}(a)--(b) visualizes the expert-load
distribution in every layer. Each row represents one layer, each column one
logical expert, and darker cells indicate that the expert receives more tokens
in that layer. Within most rows, a small subset of experts receives several
times the load expected under uniform routing, while the remaining experts
receive substantially less.
Figure~\ref{fig:expert-load-hotspots}(c) quantifies this skew. The
maximum-to-mean ratio averages $7.34$ on DAPO-Math and $7.08$ on
StarCoderData, remains above five in nearly every layer, and peaks at $12.93$
and $8.85$, respectively. Under synchronized EP execution, such a hotspot can
increase its host rank's workload and delay the entire layer.
Figure~\ref{fig:expert-load-hotspots}(d) further shows that hotspot identities
change across adjacent microbatches: averaged across layers and microbatch
pairs, $15.9\%$ and $20.0\%$ of the eight highest-load experts change on
DAPO-Math and StarCoderData, respectively. Together, these results show that
expert hotspots are common and vary across both layers and microbatches.
Consequently, a fixed expert placement can become stale as routing loads
change~\cite{hwang2023tutel,he2022fastermoe}.

Under a fixed expert-to-rank mapping, expert-level skew can translate into
rank-level imbalance. Ranks hosting hot experts process more tokens and become
stragglers, while other ranks finish earlier and wait at synchronization
points~\cite{he2022fastermoe,10.1145/3503221.3508417}. The same skew also creates uneven
communication~\cite{li2023lina,liu2023janus}: hot ranks receive more token
activations during dispatch and return more outputs during combine.
Consequently, they can become stragglers in both computation and communication.

\subsection{Topology-Dependent Cost}\label{subsec:heterogeneous-network}

Equal maximum rank loads do not imply equal communication costs.
Scale-out EP groups span NVLink domains interconnected by RDMA,
so replica placement determines which token, parameter, and
replica-gradient transfers cross domains.

\begin{figure}[!htbp]
    \centering
    \includegraphics[width=\columnwidth]{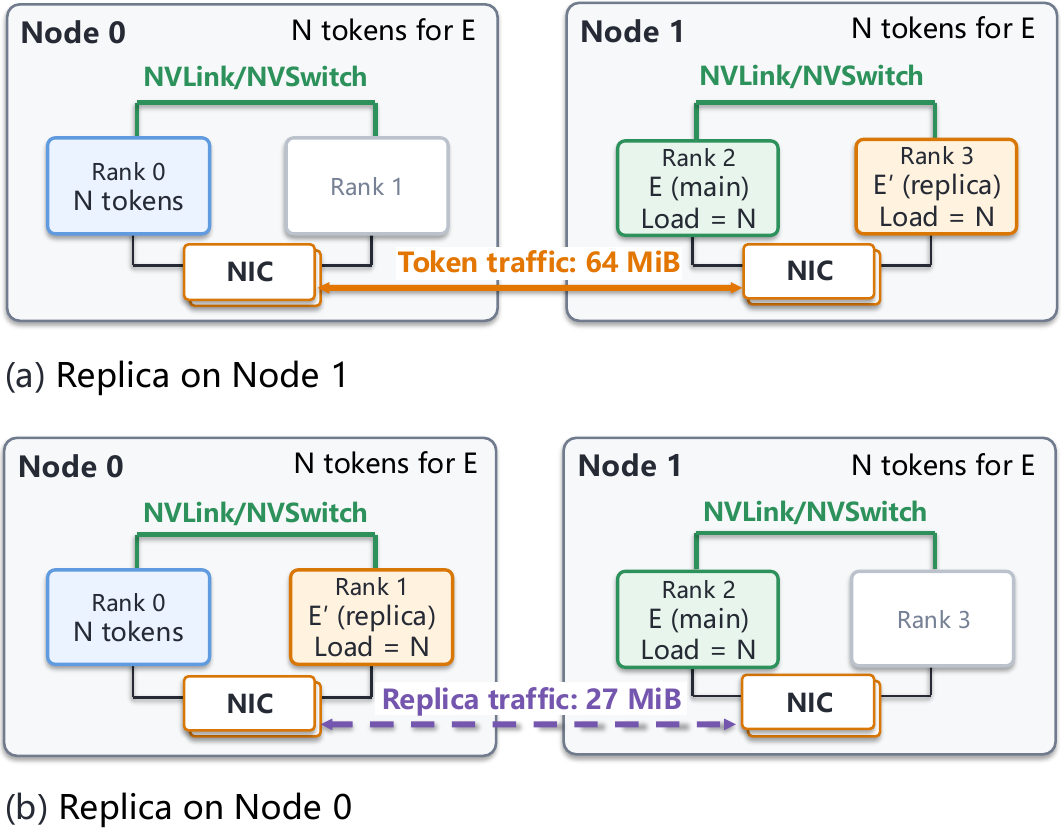}
    \caption{
    Inter-node communication for two replica placements.
    Each node contributes $N=4096$ tokens to expert $E$.
    Both placements use one replica and have the same maximum
    per-rank load, $L_{\max}=N$.
    (a) A replica on Node~1 incurs 64~MiB of token traffic
    (solid arrow).
    (b) A replica on Node~0 incurs 18~MiB of parameter traffic
    and 9~MiB of replica-gradient traffic (dashed arrow).
    }
    \label{fig:placement-cost}
\end{figure}

Figure~\ref{fig:placement-cost} illustrates this tradeoff with
two nodes, each forming an NVLink domain. Each node contributes
$N=4096$ tokens routed to expert $E$, whose main instance resides
on Node~1. Both placements use one replica, $E'$, and assign
$N$ tokens to each instance, giving the same maximum per-rank
load, $L_{\max}=N$. The example uses Qwen3-30B-A3B's
2,048-dimensional activations and 9~MiB of parameters per expert,
with BF16 activations, parameters, and replica gradients.

In Figure~\ref{fig:placement-cost}(a), inter-node token traffic
totals 64~MiB over the forward and backward passes.
Figure~\ref{fig:placement-cost}(b) keeps token routes local,
replacing this traffic with 18~MiB of parameter traffic and
9~MiB of replica-gradient traffic. Two parameter fetches are required
because replica buffers are reused across layers.

The second placement therefore reduces RDMA traffic by a factor
of approximately 2.4 while preserving the maximum per-rank load.
This example motivates jointly accounting for rank load,
token routing, and replica-transfer cost.

\subsection{Planning on the Critical Path}
\label{subsec:online-eplb-critical-path}

An online EPLB planner that adapts to the current microbatch depends
on the routing decisions produced by Top-$k$ gating. Its replica-placement and token-routing decisions
must then be available before the dependent dispatch operations can
begin. This dependency places planning on the execution path from
gating to dispatch.

\parab{Host-side planning.}
When the planner runs on the CPU, device-resident load statistics
must be made available to the host, and the resulting plan must be
returned to the GPUs. For a serialized implementation, the latency
from device-input readiness to device-plan readiness can be
decomposed as
\begin{equation}
    T_{\mathrm{host\text{-}plan}}
    = T_{\mathrm{sync}}
    + T_{\mathrm{CPU\text{-}solve}}
    + T_{\mathrm{return}},
    \label{eq:planning-overhead}
\end{equation}
where $T_{\mathrm{sync}}$ includes synchronization and the D2H input transfer,
$T_{\mathrm{CPU\text{-}solve}}$ is the CPU solving time, and
$T_{\mathrm{return}}$ is the H2D result transfer. These serialized stages block
dependent GPU work until the plan becomes available.

\begin{figure}[t]
    \centering
    \includegraphics[width=\columnwidth]{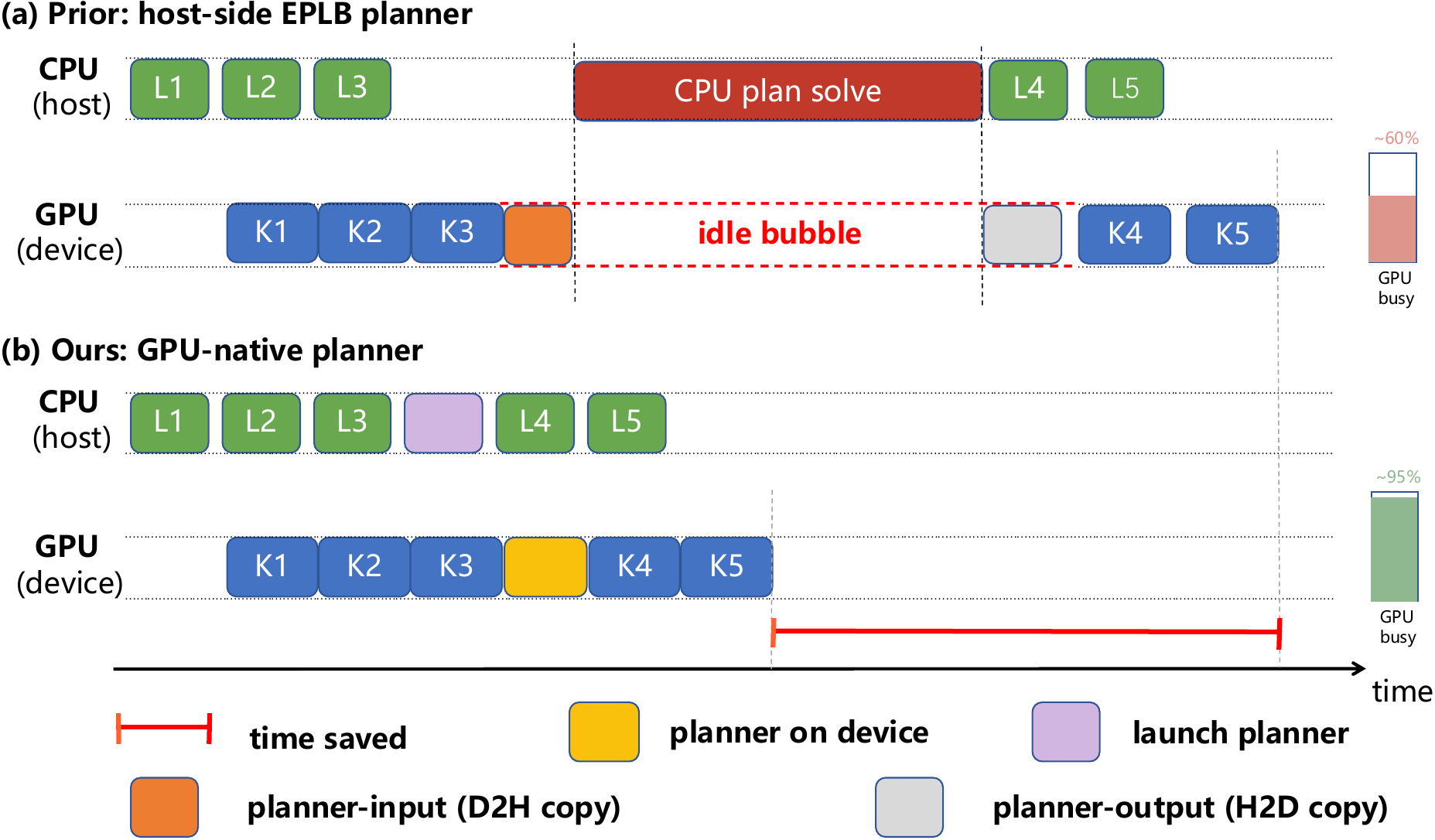}
    \caption{Critical-path execution of host-side and GPU-native EPLB
    planning. GPU-native planning removes the host round trip between
    routing and dependent dispatch operations.}
    \label{fig:planner-critical-path}
\end{figure}

As illustrated in Figure~\ref{fig:planner-critical-path},
host-side planning causes a data-dependent GPU--CPU--GPU
round trip between gating and dispatch. Although asynchronous
kernel submission can overlap host execution with GPU work,
dependent dispatch operations must wait for CPU solving and
plan transfer to complete. When insufficient independent GPU
work is available, this dependency exposes a GPU idle interval.
For a fixed workload and GPU configuration, such stalls
lengthen the training step and reduce model FLOPs utilization (MFU).
These delays can recur at every MoE layer and microbatch,
increasing their impact on training efficiency.
GPU-native planning removes the host round trip by generating
plans on the device for direct consumption by subsequent GPU
operations, eliminating this source of exposed control latency.

\parab{End-to-end benefit.}
Dynamic EPLB reduces training time only when the saved expert FFN computation
exceeds the additional planning and expert-transfer overhead. Let
$T_{\mathrm{saved}}$ denote the expert-computation time saved relative to
static placement, $T_{\mathrm{plan}}$ the GPU solver latency, and
$T_{\mathrm{transfer}}$ the exposed cost of transferring expert parameters and
replica gradients. The net-benefit condition is
\begin{equation}
    T_{\mathrm{saved}}
    > T_{\mathrm{plan}} + T_{\mathrm{transfer}}.
    \label{eq:eplb-net-benefit}
\end{equation}
The GPU solver must therefore generate each plan with low latency, while the
execution path minimizes or overlaps parameter and gradient transfers. These
requirements motivate the GPU-native design of \textit{TopoEP}, described
in Section~\ref{sec:system-design}.

\section{System Design}\label{sec:system-design}
\begin{figure}[t]
    \centering
    \includegraphics[width=\columnwidth]{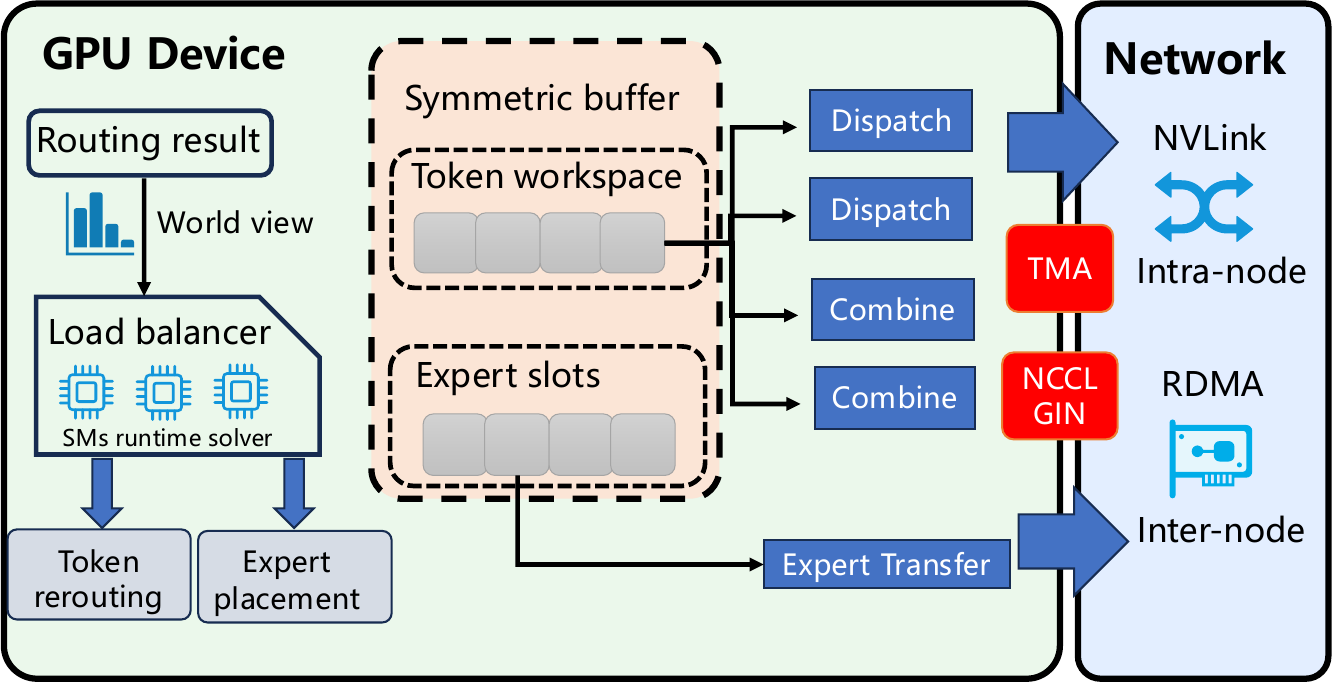}
    \caption{Architecture and execution flow of \textit{TopoEP}.}
    \label{fig:system-overview}
\end{figure}

Figure~\ref{fig:system-overview} summarizes the control and data flows of
\textit{TopoEP}. The globally gathered routing information serves as input
to the load balancer, which generates the replica placement and token
routing (Section~\ref{subsec:load-balancer}). The replica placement drives
expert-parameter transfers into reusable replica slots, using TMA over
intra-node NVLink and NCCL GIN over inter-node RDMA
(Section~\ref{subsec:buffer-management}). The token-routing decisions
drive dispatch and combine through the token workspace, while the two-chunk
pipeline overlaps communication with expert computation
(Section~\ref{subsec:comm-comp-overlap}). Together, these stages keep both
planning and execution on the GPUs.

\subsection{GPU-Native Load Balancer}
\label{subsec:load-balancer}

After Top-$k$ gating, \textit{TopoEP} first performs an EP-group all-gather
to collect token-routing information from all ranks, giving every rank the same
global routing matrix $\Omega$. The load balancer consumes $\Omega$ and
generates the replica placement and token routing for the current MoE layer
and microbatch. Given the same routing matrix, all ranks independently derive
the same plan without a coordinator or result broadcast.

The placement and routing decisions remain as device tensors and directly
drive the replica-transfer and token-dispatch paths shown in
Figure~\ref{fig:system-overview}. All inputs, intermediate states, and outputs
remain in GPU memory, avoiding data-dependent host synchronization between
planning and execution. Section~\ref{sec:algorithm} details the placement and
routing algorithm and its parallel GPU implementation.

\subsection{Replica Buffer Management}
\label{subsec:buffer-management}

\textit{TopoEP} keeps one main instance of each logical expert $e$ on rank
$\operatorname{main}(e)$, where its model state, optimizer state, and checkpoint
ownership remain throughout training. Taking unsharded BF16 mixed-precision
training with Adam as an example, a BF16 parameter value, a BF16 gradient, an
FP32 master copy, and two FP32 moment estimates together occupy 16 bytes per
parameter. A temporary replica receives only the BF16 parameters needed for
expert computation and returns its gradients to the main rank for the optimizer
update. The optimizer state and checkpoint ownership therefore never migrate
with replica placement.

Each rank preallocates a fixed number of replica slots and reuses them across
layers and microbatches. Applying a new placement only updates the expert-to-slot
mapping and copies the selected BF16 parameters. During backward propagation,
the same storage is repurposed for replica gradients after the parameters have
been consumed. An occupied slot therefore requires only 2 bytes per parameter
at any time, one eighth of the complete Adam training state.

\parab{Device-initiated replica transfers.}

\begin{figure}[t]
    \centering
    \includegraphics[width=\columnwidth]{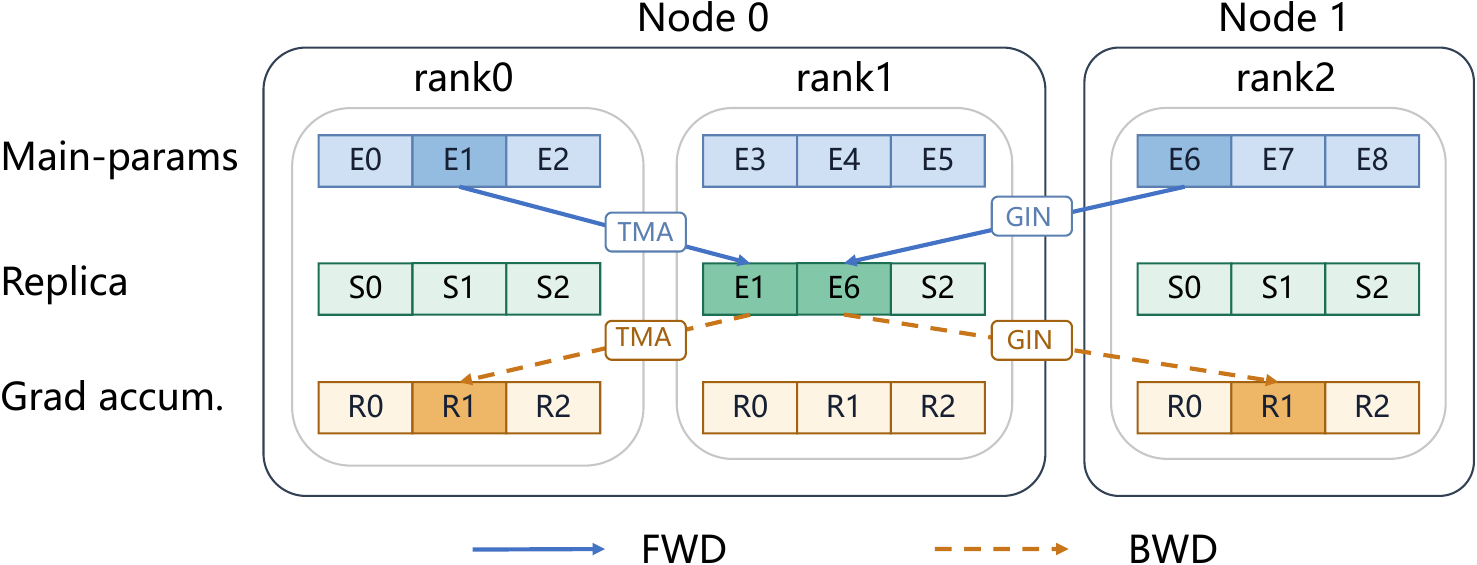}
    \caption{Device-initiated replica transfer using intra-node TMA and inter-node NCCL GIN: forward pulls expert parameters into replica slots, and backward returns replica gradients to their main ranks.}
    \label{fig:gin-replica-transfer}
\end{figure}

At runtime, the generated placement identifies the main rank associated with
each occupied replica slot and therefore determines both the parameter source
and gradient destination. As shown in
Figure~\ref{fig:gin-replica-transfer}, the CUDA kernels in
\textit{TopoEP} consume this mapping directly: the forward pass transfers
expert parameters from main instances to replica slots, and the backward pass
returns replica gradients to the gradient-accumulation buffers on the
corresponding main ranks. Initiating these transfers directly from the GPU
avoids copying the schedule to the CPU and introducing host synchronization
between planning and execution.

\textit{TopoEP} selects the device-initiated transfer mechanism according
to the network path. Tensor Memory Accelerator (TMA)~\cite{nvidia_hopper_2022}
handles intra-node transfers over NVLink, while NCCL GPU-Initiated Networking
(GIN)~\cite{hamidouche2025gpu} provides one-sided inter-node RDMA over
InfiniBand or RoCE. GIN exposes registered communication buffers through
symmetric-memory windows. After the buffers are collectively registered during
initialization, a GPU kernel addresses remote memory using a window handle,
peer rank, and window-relative offset.

Assume that the main instances of the $E$ logical experts are evenly
distributed across $P$ EP ranks and that each expert's parameters occupy
$\lvert W\rvert$ bytes. For $N_{\mathrm{slot}}$ replica slots per rank, the
three communication buffers have the following per-rank capacities and are
reused throughout training:

\begin{itemize}
    \item \textbf{Main-parameter buffer}, with a size of
    $\frac{E}{P} \cdot \lvert W\rvert$ bytes, makes the parameters of main
    instances owned by the current rank accessible to ranks hosting their
    replicas.

    \item \textbf{Replica buffer}, with a size of
    $N_{\mathrm{slot}} \cdot \lvert W\rvert$ bytes, stores the parameters of
    replicas instantiated on the current rank and is reused to stage and send
    their gradients during backward propagation.

    \item \textbf{Gradient-accumulation buffer}, with a size of
    $E \cdot \lvert W\rvert$ bytes, receives replica gradients for the current
    rank's main instances. The buffer is partitioned by source
    rank into $P$ regions, each containing
    $\frac{E}{P} \cdot \lvert W \rvert$ bytes.
\end{itemize}

\subsection{Fine-Grained Pipelined Execution}
\label{subsec:comm-comp-overlap}

A standard expert-parallel MoE layer communicates tokens during dispatch and combine. Dynamic replication additionally incurs GPU solver execution, expert-parameter transfers, and replica-gradient transfers. Executing these operations sequentially would place their aggregate latency on the MoE critical path. \textit{TopoEP} instead partitions the routed tokens into two chunks and schedules communication and expert FFN computation on separate CUDA streams. This pipeline overlaps communication for one chunk with computation for the other, reducing the exposed component of $T_{\mathrm{transfer}}$.

\subsubsection{Forward Pass}

\begin{figure}[t]
    \centering
    \includegraphics[width=\columnwidth]{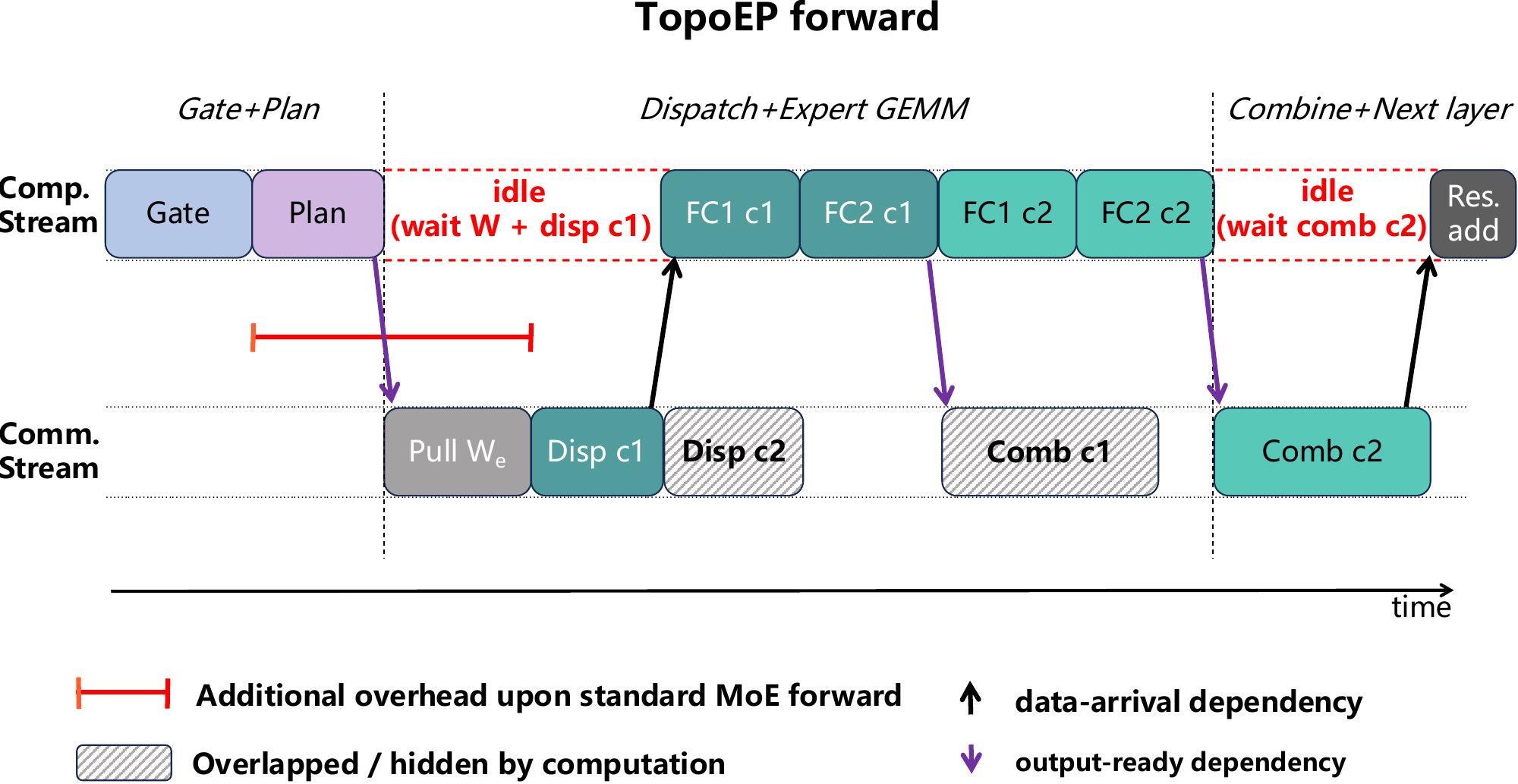}
    \caption{Two-chunk forward pipeline overlapping expert-parameter transfers and token communication with expert FFN computation.}
    \label{fig:forward-pipeline}
\end{figure}

After the global token-routing all-gather and GPU planning, each rank transfers the parameters required by its expert-to-slot mapping into the corresponding replica slots. The routed tokens are then partitioned into two disjoint chunks, $c_1$ and $c_2$. As shown in Figure~\ref{fig:forward-pipeline}, the communication stream dispatches $c_1$ before its expert FFN computation begins. While the compute stream processes $c_1$, the communication stream dispatches $c_2$. It subsequently combines the outputs of $c_1$ while the compute stream processes $c_2$.

This schedule hides the dispatch of $c_2$ and the combine of $c_1$ behind expert FFN computation. The dispatch of $c_1$ and the combine of $c_2$ remain as the pipeline fill and drain boundaries. Compared with the original expert-parallel forward path, the added critical-path latency arises primarily from the GPU solver kernels and expert-parameter transfers. These costs correspond to $T_{\mathrm{plan}}$ and the forward component of $T_{\mathrm{transfer}}$, respectively.

\subsubsection{Backward Pass}

\begin{figure}[t]
    \centering
    \includegraphics[width=\columnwidth]{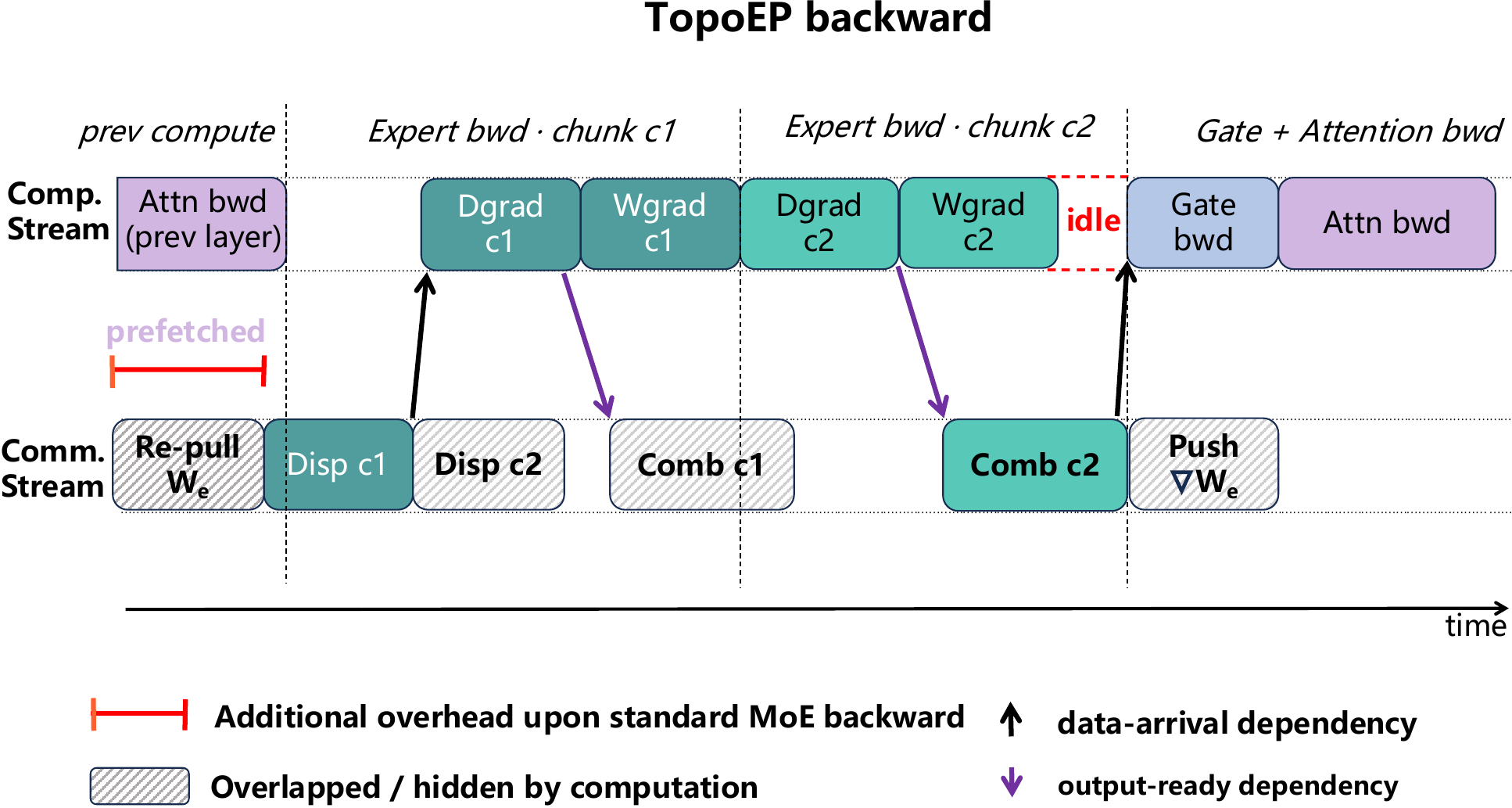}
    \caption{Two-chunk backward pipeline overlapping expert-parameter re-pulls, replica-gradient pushes, and token communication with expert FFN backward computation.}
    \label{fig:backward-pipeline}
\end{figure}

Replica slots are reused by subsequent MoE layers during forward propagation,
so a layer's temporary expert parameters are not retained until its backward
pass. \textit{TopoEP} caches the per-layer replica placement and slot
mapping, then reloads the required BF16 parameters from the main rank's
main-parameter buffer into the local replica buffer. The communication stream
issues this reload before the parameters are consumed and overlaps it with
available computation.

The backward pass of each expert GEMM comprises a data-gradient (Dgrad) GEMM and
a weight-gradient (Wgrad) GEMM. Dgrad reads the corresponding weight matrix to
propagate gradients toward the expert input, whereas Wgrad depends only on the
saved activations and output gradients. \textit{TopoEP} schedules each
chunk's Dgrad before its Wgrad, allowing the resulting token gradients to enter
the backward communication path without waiting for weight-gradient
computation.

The two chunks follow the schedule in Figure~\ref{fig:backward-pipeline}. While
the compute stream executes the expert backward kernels for $c_1$, the
communication stream dispatches the token gradients of $c_2$. The combine
operation for $c_1$ then overlaps the expert backward computation of $c_2$.
The prefetched expert parameters are temporarily staged outside the replica
buffer, so the freed buffer is reused to accumulate the Wgrad contributions
from $c_1$ and $c_2$ before transferring the result to the main rank's
gradient-accumulation buffer. An
EP-group-wide fence ensures that all remote transfers complete before the main
rank sums these regions and returns the resulting gradients to autograd. The
returned tensors do not alias the accumulation buffer, allowing it to be
cleared and reused after the preceding stream operations complete.

With a sufficient overlap window, expert-parameter reload and replica-gradient
transfer can be hidden by concurrent computation. The exposed incremental
overhead then lies primarily in the forward pass, where the GPU solver
generates the plan and transfers the selected expert parameters into replica
slots. These costs correspond to $T_{\mathrm{plan}}$ and the forward component
of $T_{\mathrm{transfer}}$, and determine whether the reduction in expert FFN
computation time satisfies the end-to-end performance criterion in
Eq.~\ref{eq:eplb-net-benefit}.

\section{Topology-Aware Load-Balancing Algorithm}\label{sec:algorithm}
\subsection{System Model}

We consider an expert-parallel MoE training cluster with $M$ NVLink domains. Let
$\mathcal{D}=\{0,\ldots,M-1\}$, $\mathcal{R}=\{0,\ldots,R-1\}$, and
$\mathcal{E}=\{0,\ldots,E-1\}$ denote the sets of domains, EP ranks, and logical
experts, respectively. The function $\operatorname{dom}(r)\in\mathcal{D}$ maps
rank $r$ to its NVLink domain. We represent the network by a symmetric
communication-time matrix $C=[c_{s,r}]$, where $c_{r,r}=0$ and
$c_{s,r}$ denotes the effective time to transfer one byte from source rank $s$
to destination rank $r$. The coefficients can be obtained from profiled effective bandwidth,
and are therefore lower for intra-domain NVLink paths than for inter-domain
RDMA paths.

Each logical expert $e$ has a fixed main instance on rank
$\operatorname{main}(e)\in\mathcal{R}$. This instance holds the expert's
parameters throughout training. Load balancing may create temporary replicas
on other ranks by copying these parameters. Let $\lVert W_e\rVert$ denote the
parameter size of expert $e$ in bytes and $S_{\mathrm{tok}}$ the size in bytes
of one routed activation or activation gradient. Each rank reserves
$N_{\mathrm{slot}}$ slots for temporary replicas in addition to its fixed main
instances.

For each microbatch, Top-$k$ gating produces the routing matrix
\begin{equation}
    \Omega=[\omega_{s,e}]
    \in\mathbb{Z}_{\geq 0}^{R\times E},
\end{equation}
where $\omega_{s,e}$ is the number of token--expert assignments originating
from source rank $s$ and selecting expert $e$. The demand for expert $e$ generated
within domain $d$ is
\begin{equation}
    D_{d,e} =
    \sum_{\substack{s\in\mathcal{R}\\
    \operatorname{dom}(s)=d}}\omega_{s,e}.
\end{equation}
Since all experts in a layer share the same FFN architecture and have similar
per-token compute costs, we estimate each rank's compute time by multiplying
its assigned token count by the profiled average per-assignment time
$t_{\mathrm{FFN}}$.

\subsection{Load-Balancing Formulation}

Given $\Omega$, the optimization jointly determines temporary replica
placement and the distribution of each source--expert demand among physical
instances of the same logical expert. It preserves every token--expert
assignment while minimizing the estimated exposed time of expert computation,
token routing, and expert-parameter transfer.

\parab{Decision variables.}
We introduce two decision variables. The binary variable $x_{e,r}$ indicates
whether rank $r$ hosts an instance of expert $e$, including its main instance.
The nonnegative integer $q_{s,e,r}$ gives the number of $(s,e)$ assignments
executed on destination rank $r$.

\parab{Load metric.}
The assigned load of rank $r$ is the total number of token--expert assignments
executed on that rank. We define $L_{\max}$ as the maximum assigned load across
all ranks.
\begin{equation*}
\begin{array}{@{}l@{\;}c@{\;}l@{}}
    L_r & = &
    \displaystyle\sum_{s\in\mathcal{R}}\sum_{e\in\mathcal{E}}q_{s,e,r},
    \quad \forall r\in\mathcal{R}, \\
    L_{\max} & = & \displaystyle\max_{r\in\mathcal{R}}L_r.
\end{array}
\end{equation*}

\parab{Constraints.}
A feasible replica-placement and token-rerouting plan satisfies the following
conditions.

\noindent\textbf{Token conservation (C1).}
Every token--expert assignment is mapped to exactly one physical instance.
\begin{equation*}
    \sum_{r\in\mathcal{R}}q_{s,e,r}=\omega_{s,e},
    \quad
    \forall s\in\mathcal{R},\ e\in\mathcal{E}.
\end{equation*}

\noindent\textbf{Instance reachability (C2).}
Tokens may be assigned only to ranks that host the corresponding expert.
\begin{equation*}
    q_{s,e,r}\leq\omega_{s,e}x_{e,r},
    \quad
    \forall s,r\in\mathcal{R},\ e\in\mathcal{E}.
\end{equation*}

\noindent\textbf{Replica capacity (C3).}
Each rank may host at most $N_{\mathrm{slot}}$ non-main expert instances in its
preallocated replica slots.
\begin{equation*}
    \sum_{\substack{e\in\mathcal{E}\\
    \operatorname{main}(e)\neq r}}x_{e,r}
    \leq N_{\mathrm{slot}},
    \quad
    \forall r\in\mathcal{R}.
\end{equation*}

\noindent\textbf{Cross-domain replica criterion (C4).}
Under the two-chunk execution model in
Section~\ref{subsec:comm-comp-overlap}, approximately half of the
$4D_{d,e}S_{\mathrm{tok}}$ bytes of forward and backward remote-token traffic
remains exposed, while backward replica transfers can be hidden when the
overlap window is sufficient. Assuming a common per-byte RDMA cost, a replica
of expert $e$ may be placed in a domain different from its main instance only
if the forward parameter transfer is smaller than the exposed remote-token
traffic:
\begin{equation*}
\begin{aligned}
    &x_{e,r}=1,\quad
    \operatorname{dom}(r)\neq
    \operatorname{dom}(\operatorname{main}(e)) \\
    &\qquad\Longrightarrow\quad
    \lVert W_e\rVert
    <2D_{\operatorname{dom}(r),e}S_{\mathrm{tok}}.
\end{aligned}
\end{equation*}

\noindent\textbf{Fixed main placement (C5).}
Every expert retains its main instance throughout training.
\begin{equation*}
    x_{e,\operatorname{main}(e)}=1,
    \quad
    \forall e\in\mathcal{E}.
\end{equation*}

\parab{Objective.}
The exposed expert FFN computation time is
\begin{equation}
    T_{\mathrm{comp}} = t_{\mathrm{FFN}}L_{\max}.
\end{equation}
Under the overlap assumptions in (C4), the exposed token-routing and forward
expert-parameter transfer times are
\begin{equation}
    T_{\mathrm{token}}
    =
    2S_{\mathrm{tok}}
    \sum_{\substack{s,r\in\mathcal{R}\\e\in\mathcal{E}}}
    c_{s,r}q_{s,e,r}.
\end{equation}
\begin{equation}
    T_{\mathrm{replica}}
    =
    \sum_{e\in\mathcal{E}}
    \sum_{\substack{r\in\mathcal{R}\\
    r\neq\operatorname{main}(e)}}
    c_{\operatorname{main}(e),r}
    \lVert W_e\rVert x_{e,r}.
\end{equation}
The optimization minimizes their sum as an idealized estimate of exposed
execution time:
\begin{equation}
\begin{aligned}
    \min_{x,q}\quad
        &T_{\mathrm{comp}}
        +T_{\mathrm{token}}
        +T_{\mathrm{replica}} \\
    \text{s.t.}\quad
        &\text{(C1)--(C5)}, \\
        &x_{e,r}\in\{0,1\}, \\
        &q_{s,e,r}\in\mathbb{Z}_{\geq0},
        \quad
        \forall e\in\mathcal{E},\ s,r\in\mathcal{R}.
\end{aligned}
\label{eq:moe-replication-formulation}
\end{equation}

\subsection{Two-Stage GPU-Native Solver}
\label{subsec:two-stage-solver}

Solving Eq.~\ref{eq:moe-replication-formulation} exactly for every layer and
microbatch would add excessive latency to the training critical path.
\textit{TopoEP} therefore constructs a feasible placement-and-routing plan
with a topology-aware two-stage GPU solver. Inter-node placement first creates
replicas when the reduction in cross-domain token traffic justifies the
parameter transfer, and intra-node refinement then reduces the remaining
rank-load imbalance within each NVLink domain.

\begin{figure*}[t]
    \centering
    \includegraphics[width=\textwidth]{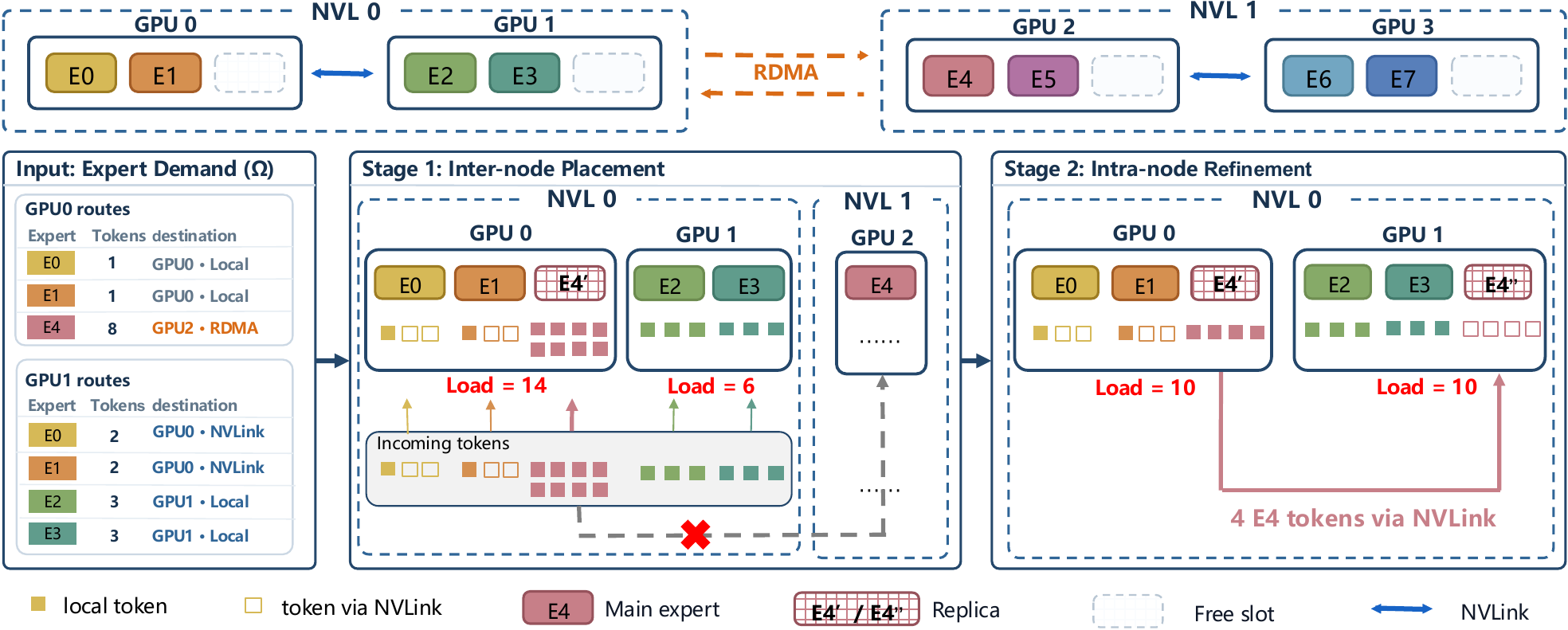}
    \caption{Example of two-stage expert replication and token rerouting.
    Inter-node placement removes cross-domain token traffic, after which
    intra-node refinement balances rank loads through same-domain rerouting
    over NVLink.}
    \label{fig:algorithm-overview}
\end{figure*}

Figure~\ref{fig:algorithm-overview} illustrates the two stages using two
NVLink domains. In the input routing, eight assignments from GPU~0 select expert
$E_4$, whose main instance resides on GPU~2. These assignments therefore cross
the inter-node RDMA fabric during both dispatch and combine. Inter-node
placement creates replica $E_4'$ on GPU~0 when one parameter transfer costs
less than the exposed bidirectional token traffic. The eight assignments can
then execute locally, eliminating their RDMA token traffic. This placement,
however, leaves GPU~0 and GPU~1 with loads of 14 and 6. Intra-node refinement
creates replica $E_4''$ on GPU~1 and moves four $E_4$ assignments from GPU~0
to GPU~1 over NVLink, balancing both rank loads at 10 without introducing new
cross-domain token routes.

\parab{Inter-node placement.}
For every domain $d$, the solver aggregates the routing matrix by source
domain and computes the replication benefit
\begin{equation}
    b_{d,e}=2D_{d,e}S_{\mathrm{tok}}-\lVert W_e\rVert.
\end{equation}
For an expert whose main instance lies outside domain $d$, a positive
$b_{d,e}$ means that replacing the exposed remote-token traffic with one
parameter transfer is beneficial. Each domain considers its positive-benefit
candidates in descending order and places replicas on the least occupied ranks
with free slots. Processing domains independently allows these decisions to
run concurrently while enforcing the per-rank slot limit.

Given the resulting placement, the solver constructs the physical token
routing. For each source--expert pair $(s,e)$, it selects instances in the
source domain whenever they are available and otherwise uses all deployed
instances of $e$. It divides $\omega_{s,e}$ approximately evenly among the
selected instances. It then computes the per-instance loads
$U_{e,r}=\sum_s q_{s,e,r}$ and updates the rank loads $L_r$. This routing
preserves every logical token--expert assignment while preferring same-domain
destinations.

\parab{Intra-node refinement.}
Starting from the inter-node placement, each NVLink domain independently
reduces its residual rank-load imbalance. In every iteration, the solver
selects the busiest rank $r_b$ and evaluates same-domain target ranks $r_t$ for
the experts contributing to its load. A target is feasible if it already hosts
the expert or has a free replica slot. The candidate transfer amount is
\begin{equation}
    \delta=\min\!\left(
        U_{e,r_b},
        \left\lfloor\frac{L_{r_b}-L_{r_t}}{2}\right\rfloor
    \right).
\end{equation}
The half-gap bound reduces the difference between the source and target loads
without reversing their order. Candidates are ranked by transfer amount,
target load, and communication cost. The selected update creates a replica
when needed and modifies $x$, $q$, $U$, and $L$. Because the target lies in the
same domain as the overloaded rank, refinement improves load balance without
placing the expert in a new domain or adding cross-domain token routes.

\parab{Constraint preservation.}
Inter-node placement starts from the fixed main instances, creates replicas
only on ranks with free slots, and admits a cross-domain expert--domain pair
only when $b_{d,e}>0$. It therefore satisfies replica capacity (C3), the
cross-domain replica criterion (C4), and fixed main placement (C5). Routing
partitions every $\omega_{s,e}$ completely among deployed instances, satisfying
token conservation (C1) and instance reachability (C2). Intra-node refinement
only transfers existing assignments to an existing instance or a new replica
in a free slot, and adds replicas only within a domain that already hosts the
expert. Each refinement step therefore preserves (C1)--(C5).

\parab{GPU parallelism and determinism.}
Inter-node candidate evaluation and source--expert routing run concurrently
across CUDA thread blocks. During intra-node refinement, one block handles each
NVLink domain, caches its placement and load state in shared memory, and uses
warp- and block-level reductions to select a candidate. The block commits one
update per iteration to avoid conflicting modifications. Deterministic
candidate ordering allows every rank to generate the same $x$ and $q$ from the
shared routing matrix, eliminating the need for a coordinator or plan
broadcast.

\section{Implementation}\label{sec:impl}
We integrate \textit{TopoEP} into
Megatron-LM~\cite{shoeybi2020megatronlmtrainingmultibillionparameter}. The
implementation comprises approximately 10,000 lines of Python and CUDA/C++.

\parab{GPU-native solver.}
Inter-node placement, routing update, and intra-node refinement are implemented
as CUDA kernels. Independent domains and source--expert tasks execute
concurrently, with warp- and block-level reductions used for candidate
selection.

\parab{Device-initiated replica transfers.}
Main-parameter, replica-slot, and gradient-accumulation buffers are allocated
and registered in NCCL symmetric-memory windows during initialization. TMA
handles intra-node transfers over NVLink, while NCCL GIN performs
device-initiated get/put operations for inter-node RDMA.

\parab{DeepEP dispatch and combine.}
We integrate DeepEP V2~\cite{deepep2025} (commit \texttt{af9a040}) into our
communication manager to replace Megatron-LM's NCCL-based all-to-all path. The
manager consumes device-resident routing information and derives receive
counts on the GPU, avoiding CPU synchronization during dispatch and combine.

\parab{Fine-grained two-chunk pipeline.}
We use a custom autograd function to explicitly schedule the expert computation
and communication of each MoE layer across separate CUDA streams. The pipeline
overlaps token communication and replica transfers with expert computation,
reducing their exposed critical-path overhead.

\section{Performance Evaluation}\label{sec:eval}
\subsection{Experimental Setup}
\label{subsec:setup}

\parab{Testbed.}
Our experiments run on a four-node cluster with 32 NVIDIA H800 GPUs. Each node
contains eight GPUs interconnected through NVLink and NVSwitch, providing up to
400 GB/s of aggregate bidirectional bandwidth. A rail-optimized InfiniBand
fabric provides inter-node communication, with each GPU connected to a
dual-port NVIDIA ConnectX-7 NIC through two 200 Gb/s links. The testbed supports
32-way expert parallelism and includes both intra-node and inter-node
communication paths.

\parab{Baselines.}
We compare \textit{TopoEP} with
Megatron-LM~\cite{shoeybi2020megatronlmtrainingmultibillionparameter} and three
representative MoE load-balancing methods:
DeepSeek-EPLB~\cite{deepseekai2025eplb}, the official expert-replication and
placement heuristic;
FasterMoE~\cite{he2022fastermoe}, which uses dynamic expert shadowing; and
FlexMoE~\cite{nie2023flexmoe}, which expands, shrinks, and migrates virtual
experts. We adapt each method's core load-balancing algorithm to the same
Megatron-LM codebase at commit \texttt{0ff7226}. Within each comparison, all
dynamic methods use the same additional replica budget, and all runs use the
same routing configuration.

\parab{Models and Workloads.}
We evaluate \textit{TopoEP} with
Qwen3-30B-A3B~\cite{qwen3technicalreport},
GLM-4.5-Air~\cite{zeng2025glm45}, and DeepSeek-V2~\cite{deepseekv2}. As
summarized in
Table~\ref{tab:model-configs}, these models span 128--160 routed experts,
Top-$k$ values of 6 and 8, and substantially different hidden and
expert-intermediate dimensions. Qwen3 uses PP~1/EP~32, while GLM-4.5-Air and
DeepSeek-V2 use PP~2/EP~16. To fit the models within the available GPU memory,
we reduce the number of Transformer layers while retaining their original
MoE-layer dimensions and routing configurations. This setup keeps the
evaluation focused on MoE-layer performance.

\begin{table}[t]
    \centering
    \caption{MoE model configurations used in the evaluation.}
    \label{tab:model-configs}
    \footnotesize
    \setlength{\tabcolsep}{1.2pt}
    \begin{tabular*}{\columnwidth}{@{\extracolsep{\fill}}lcccccc@{}}
        \toprule
        Model &
        \makecell{MoE\\layers} &
        Experts &
        Top-$k$ &
        Hidden &
        \makecell{Interm.\\size} &
        \makecell{PP / EP} \\
        \midrule
        Qwen3-30B-A3B~\cite{qwen3technicalreport}
            & 5 & 128 & 8 & $2{,}048$ & 768 & 1/32 \\
        GLM-4.5-Air~\cite{zeng2025glm45}
            & 5 & 128 & 8 & $4{,}096$ & $1{,}408$ & 2/16 \\
        DeepSeek-V2~\cite{deepseekv2}
            & 4 & 160 & 6 & $5{,}120$ & $1{,}536$ & 2/16 \\
        \bottomrule
    \end{tabular*}
\end{table}

We construct a training corpus from six public sources: English web text from
FineWeb~\cite{penedo2024the};
Chinese web text from FineWeb2~\cite{penedo2025fineweb2pipelinescale};
diverse pretraining text from
Dolma~\cite{soldaini-etal-2024-dolma}; source code in Python, C++, Java, and
Rust from StarCoderData~\cite{li2023starcoder}; scientific literature from
peS2o~\cite{peS2o}; and mathematical reasoning prompts from
DAPO-Math-17K~\cite{yu2025dapo}. We deduplicate and truncate documents to at
most 4,096 tokens before assembling them into training sequences.

\subsection{Load-Balancing Effectiveness}
\label{subsec:load-balancing}

\begin{figure}[t]
    \centering
    \includegraphics[width=\columnwidth]{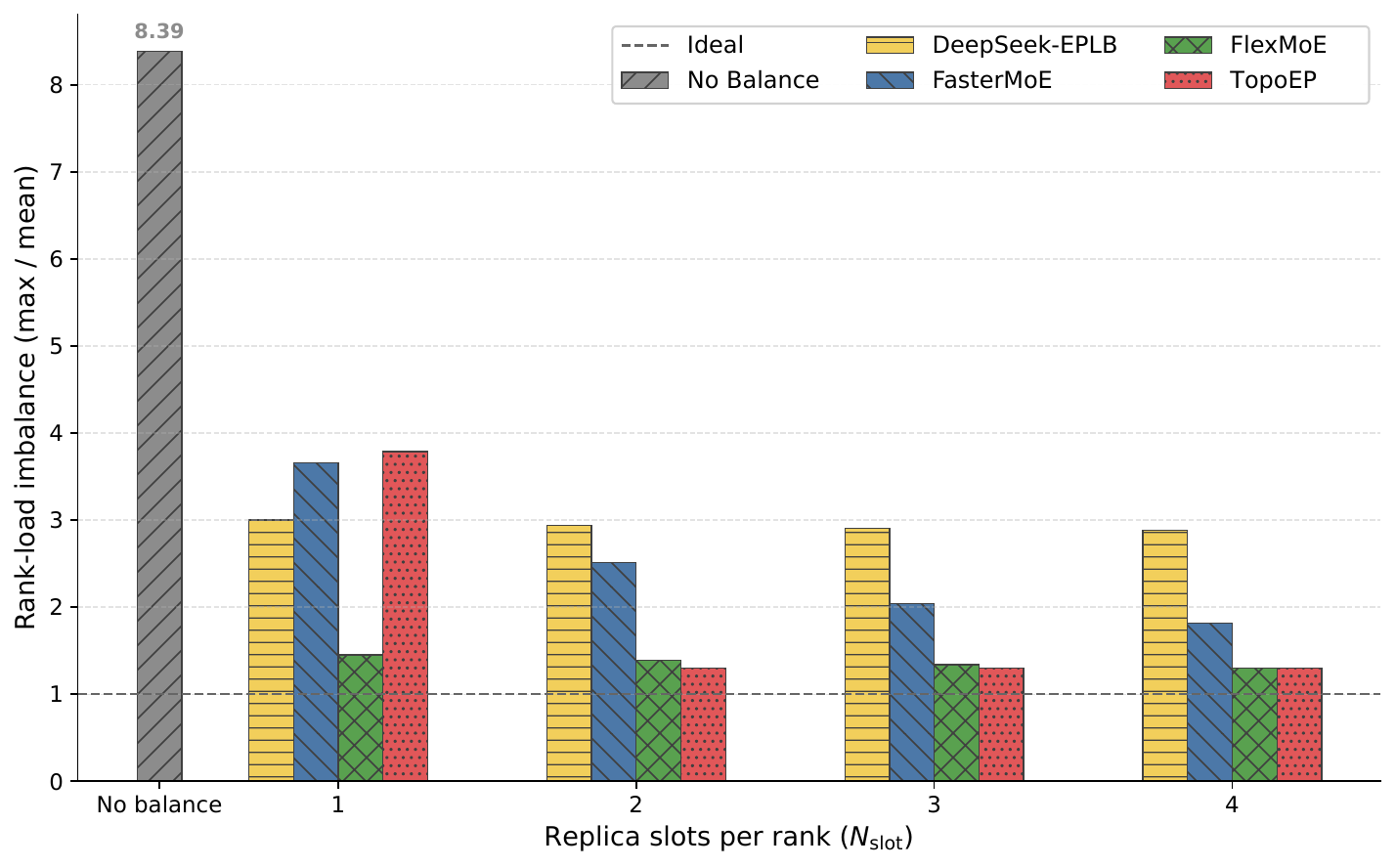}
    \caption{Rank-load imbalance versus additional replica slots per rank for
    a 32-rank, 640-expert Top-8 workload. Lower is better, and the dashed line
    denotes ideal balance.}
    \label{fig:ep-Balancing}
\end{figure}

Figure~\ref{fig:ep-Balancing} compares the maximum-to-mean rank-load ratio as
the additional replica budget increases. Without balancing, the ratio is 8.39.
With one slot, the ratio for \textit{TopoEP} remains 3.79, whereas FlexMoE
reaches 1.45. With two slots, \textit{TopoEP} drops to 1.30 and outperforms
FlexMoE, FasterMoE, and DeepSeek-EPLB by 6.8\%, 48.2\%, and 55.8\%,
respectively. With three or four slots, \textit{TopoEP} maintains a
rank-load ratio of 1.30, while FasterMoE and DeepSeek-EPLB remain at 1.82 and
2.88 even with four slots. Thus, \textit{TopoEP} needs only two additional
slots per rank to achieve the best load balance among the evaluated methods;
allocating more slots provides almost no further benefit. This small replica
budget also reduces the additional GPU memory overhead introduced by expert
replication. We use this setting in the remaining experiments.

\subsection{Solver Scalability}
\label{subsec:solver-scalability}

\begin{figure}[t]
    \centering
    \includegraphics[width=\columnwidth]{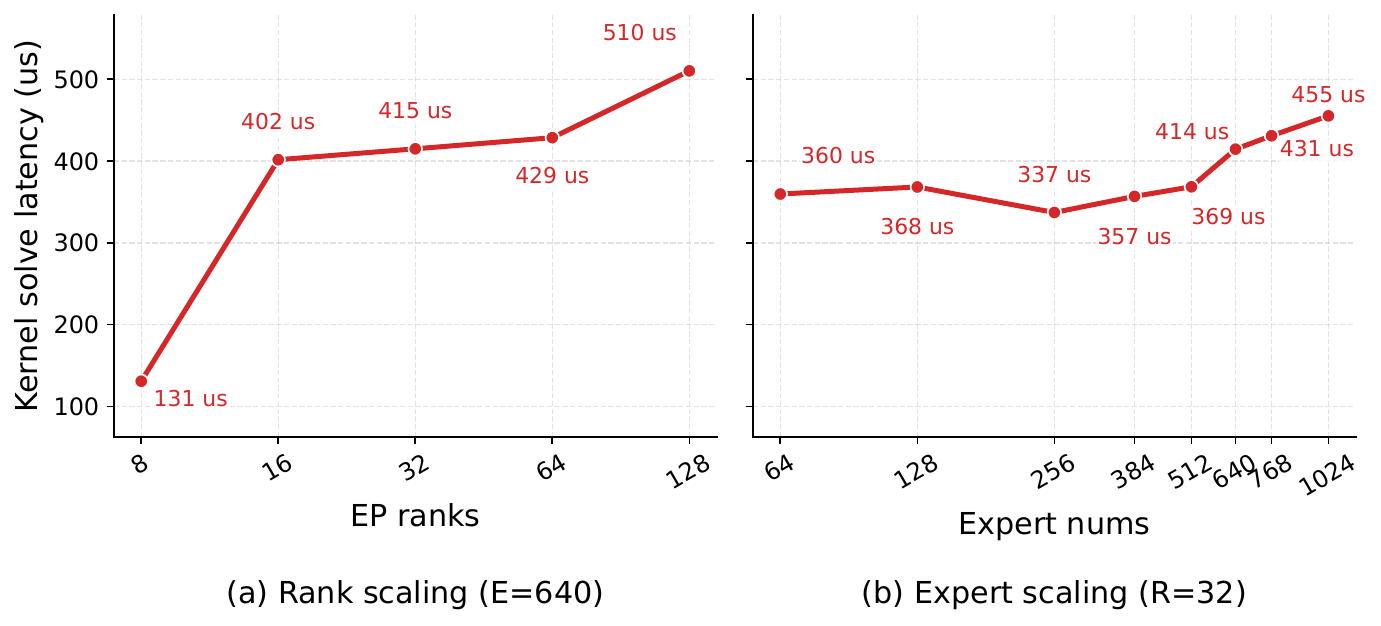}
    \caption{Mean solver-kernel latency with varying logical EP sizes and
    expert counts, measured over 200 executions after 20 warm-up iterations.}
    \label{fig:ep-scale}
\end{figure}

Figure~\ref{fig:ep-scale}(a) shows that, with the number of experts fixed at
\(E=640\), kernel latency increases
from \(131\,\mu\mathrm{s}\) at EP size 8 to \(510\,\mu\mathrm{s}\) at EP size
128. At EP size 8, all ranks fit within a single NVLink domain, so the solver
does not need to evaluate inter-node placement, contributing to the lower
kernel latency. Figure~\ref{fig:ep-scale}(b) varies the expert count by
\(16\times\) at a fixed EP size of 32, while latency remains between \(337\)
and \(455\,\mu\mathrm{s}\). Increasing the EP size by \(16\times\) raises
latency by only \(3.9\times\), while the same increase in expert count changes
the endpoint latency by only \(1.27\times\). Thus, the problem grows by an
order of magnitude without a proportional latency increase. Even the largest
tested configuration is solved in just \(0.510\,\mathrm{ms}\), preserving
microsecond-scale planning at every layer and microbatch.

\subsection{End-to-End Performance}
\label{subsec:e2e-performance}

In this section, we compare the end-to-end training performance of
\textit{TopoEP} against multiple baselines and examine how routing skew
affects training throughput.

\begin{figure*}[t]
    \centering
    \includegraphics[width=\textwidth]{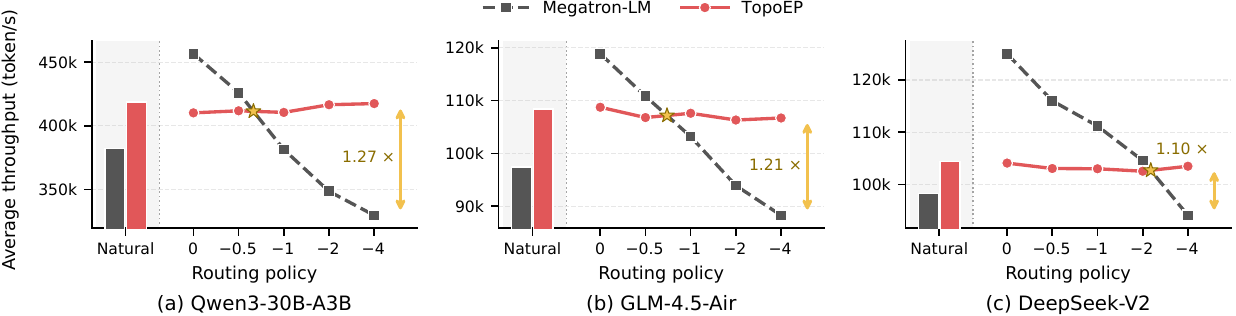}
    \caption{Average end-to-end training throughput under natural and synthetic
    routing. The synthetic settings vary \texttt{router\_skew} from 0 to
    $-4$. Stars mark interpolated throughput crossovers, and double-headed
    arrows show the \textit{TopoEP}/Megatron-LM throughput ratio at
    \texttt{router\_skew}$=-4$.}
    \label{fig:e2e-skew}
\end{figure*}

\parab{Sensitivity to routing skew.}
Natural routing uses the router logits generated by the model from the input
tokens. As discussed in Section~\ref{subsec:dynamic-load-imbalance},
expert-load imbalance commonly arises in MoE training and gives load balancing
greater opportunity to reduce straggler delays. Because expert specialization
typically emerges over long pretraining runs, our reduced-depth models and
limited training duration may not exhibit the full range of routing imbalance
that can arise at full scale. We therefore construct synthetic routing
workloads with controlled skew to test whether \textit{TopoEP} mitigates
the resulting throughput degradation across a broader range of imbalance.
Synthetic routing uses random logits with a fixed
expert-specific bias. A \texttt{router\_skew} value of 0
produces uniform expert selection in expectation, while increasingly negative
values strengthen the bias and concentrate more tokens on a subset of experts.
We use these synthetic settings only for performance measurements.
Under natural routing, Figure~\ref{fig:e2e-skew} shows that
\textit{TopoEP} improves throughput over Megatron-LM by 9.5\%, 11.4\%,
and 6.2\% on Qwen, GLM, and DeepSeek-V2, respectively. Under uniform synthetic
routing (\texttt{router\_skew}$=0$), it instead trails Megatron-LM by
8.5\%--16.7\%. Thus, load balancing does not improve every routing workload.
When expert demand is already uniform, the limited reduction in straggling
does not offset planning and replica-management costs.

As synthetic routing becomes more imbalanced from
\texttt{router\_skew}$=0$ to $-4$, Megatron-LM throughput falls by 27.8\%,
25.8\%, and 24.7\% across the three models. In contrast,
\textit{TopoEP} throughput varies by less than 2.3\% across all synthetic
settings. At \texttt{router\_skew}$=-4$, \textit{TopoEP} outperforms
Megatron-LM by 26.6\%, 20.9\%, and 10.0\%. These results show that its
throughput benefit grows as routing imbalance increases. More importantly,
\textit{TopoEP} maintains nearly constant throughput across the full skew
range. This robustness shows that substantial changes in routing and
expert-load imbalance need not translate into large performance fluctuations,
enabling stable training performance as routing behavior evolves.

\begin{figure}[t]
    \centering
    \includegraphics[width=\columnwidth]{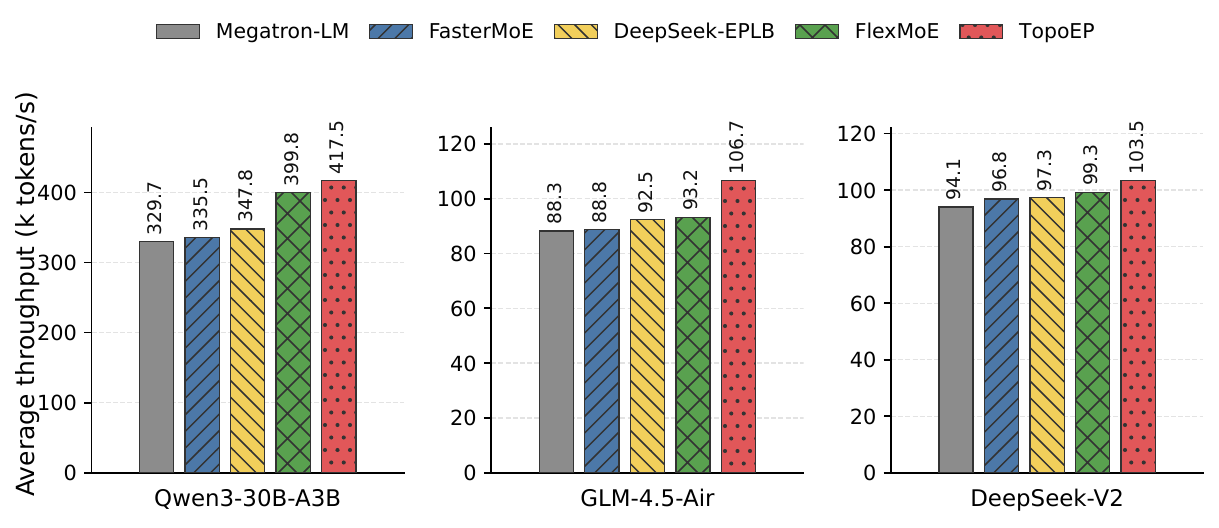}\\[-0.5ex]
    \includegraphics[width=\columnwidth]{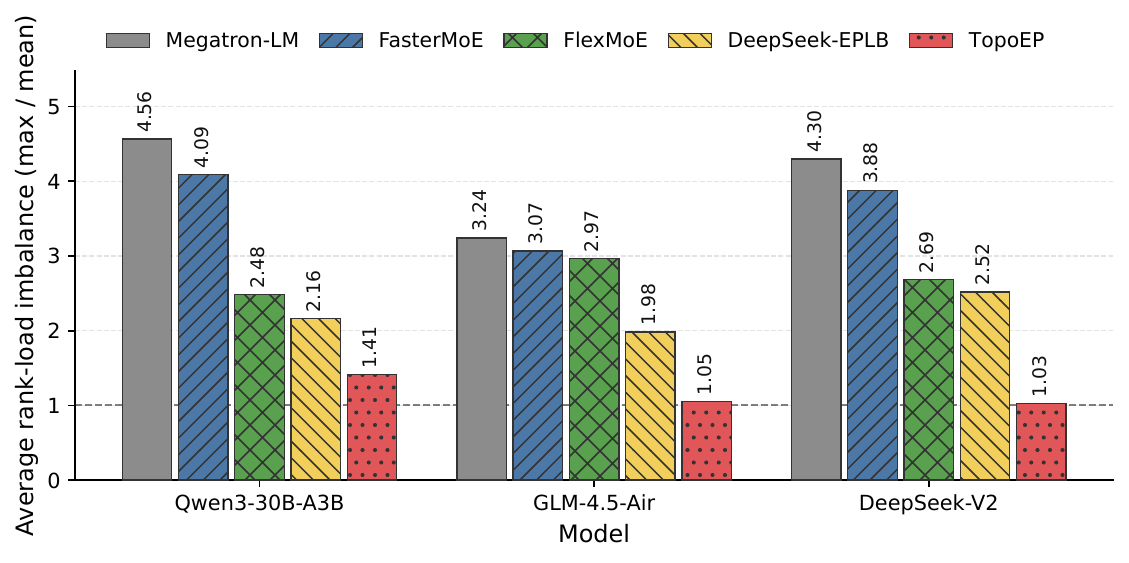}
    \caption{Comparison with baselines at
    \texttt{router\_skew}$=-4$. (a) Average end-to-end throughput. (b) Average
    rank-load imbalance, defined as the maximum per-rank token-expert load
    divided by the average across ranks, where 1 denotes ideal balance.}
    \label{fig:e2e-baselines}
\end{figure}

\parab{Comparison with baselines.}
To ensure a fair comparison of load-balancing policies, all evaluated EPLB methods
use the same optimized plan-execution backend described in
Section~\ref{sec:impl}, including TMA\,/\,GIN replica transfers, DeepEP token
dispatch and combine, and the two-chunk execution pipeline. They differ only
in the planner that generates the placement and routing decisions; Megatron-LM
serves separately as the no-EPLB baseline.

Figure~\ref{fig:e2e-baselines}(a) shows that \textit{TopoEP} achieves the
highest throughput among all evaluated methods on all three models. It exceeds
the highest baseline throughput by 4.4\% on Qwen, 14.5\% on GLM, and 4.2\% on
DeepSeek-V2. Figure~\ref{fig:e2e-baselines}(b) shows that \textit{TopoEP}
also achieves the lowest average rank-load imbalance, reducing the ratios to
1.41, 1.05, and 1.03, respectively. These values are 34.7\%, 47.1\%, and
59.2\% lower than the lowest baseline ratios.

With the plan-execution backend held constant across EPLB methods, these
results show that \textit{TopoEP} achieves better rank-load balance while
also delivering higher end-to-end throughput when planner and execution
overheads are included.

\begin{figure}[t]
    \centering
    \includegraphics[width=\columnwidth]{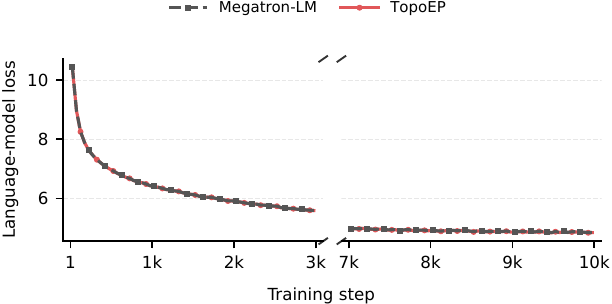}
    \caption{Qwen3-30B-A3B training loss during 10,000-step runs under natural
    routing.}
    \label{fig:loss-convergence}
\end{figure}

\parab{Training stability.}
Figure~\ref{fig:loss-convergence} shows closely overlapping loss trajectories
for \textit{TopoEP} and Megatron-LM under the same model initialization,
data order, and learning-rate schedule. Across all 10,000 steps, the mean
and maximum absolute relative differences are 0.088\% and 0.407\%,
respectively. These results show no observable degradation in loss convergence.

\subsection{Latency Breakdown}
\label{subsec:latency-breakdown}

\begin{figure}[t]
    \centering
    \includegraphics[width=\columnwidth]{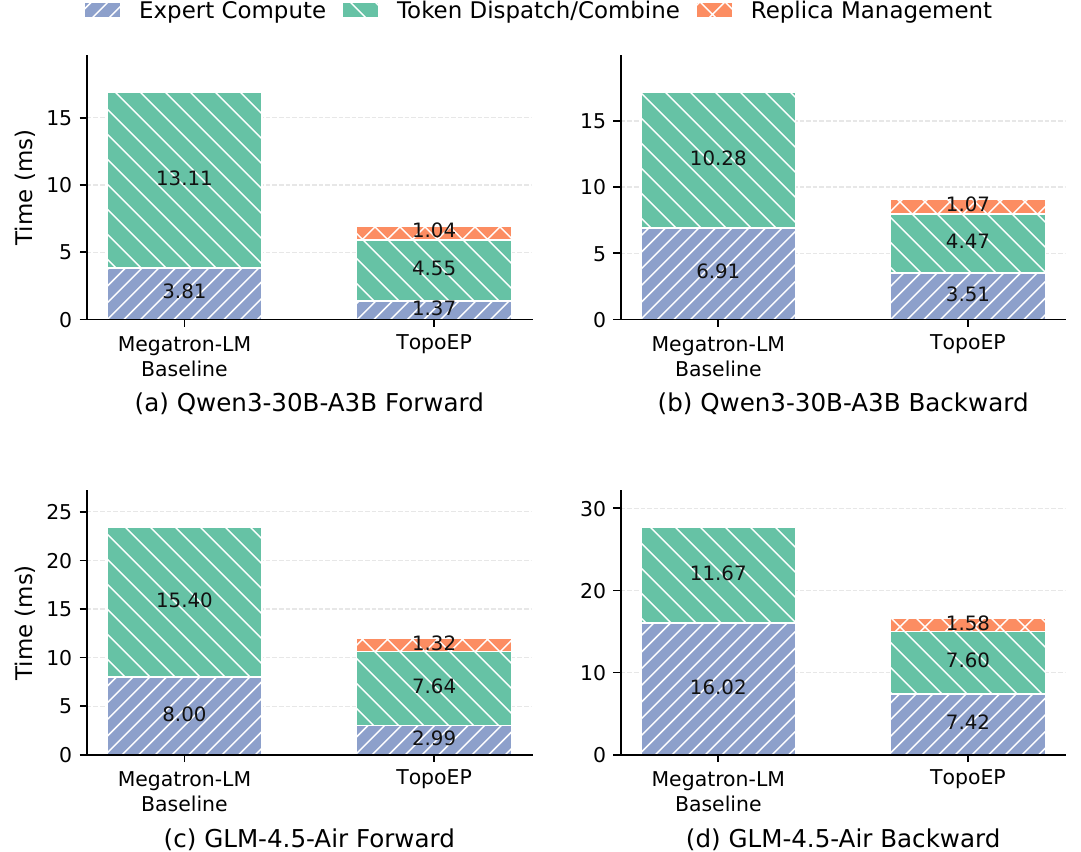}
    \caption{Per-layer MoE latency breakdown, with expert-computation time
    measured on straggler ranks.}
    \label{fig:latency-breakdown}
\end{figure}

Figure~\ref{fig:latency-breakdown} shows that \textit{TopoEP} reduces
expert-computation time on straggler ranks by 49.2\%--64.0\% across the
forward and backward passes of Qwen and GLM, directly demonstrating its
effectiveness in mitigating computation stragglers. Token dispatch/combine
time also decreases by 34.9\%--65.3\%. \textit{TopoEP} introduces
1.04--1.58\,ms of replica-management time per layer, which is smaller than the
reduction in each of the other two components in every panel. Because these
components may overlap across CUDA streams, their measured times cannot be
added directly. Overall, the breakdown shows that \textit{TopoEP} reduces
both expert-computation and token dispatch/combine time under imbalanced
routing.

\subsection{Topology-Aware Token Routing}
\label{subsec:topology-aware-routing}

We examine whether the plans generated by the algorithm in Section~\ref{sec:algorithm} are topology-aware in practice.
All methods receive the same logical Top-$k$ assignments, with a common replica budget for the EPLB methods.
They differ in the placement of expert instances and the physical ranks selected to execute each assignment.
For each token--expert assignment, we record whether its execution rank is on a different node from its source rank and whether it is served by a replica.

\begin{figure}[t]
    \centering
    \includegraphics[width=\columnwidth]
        {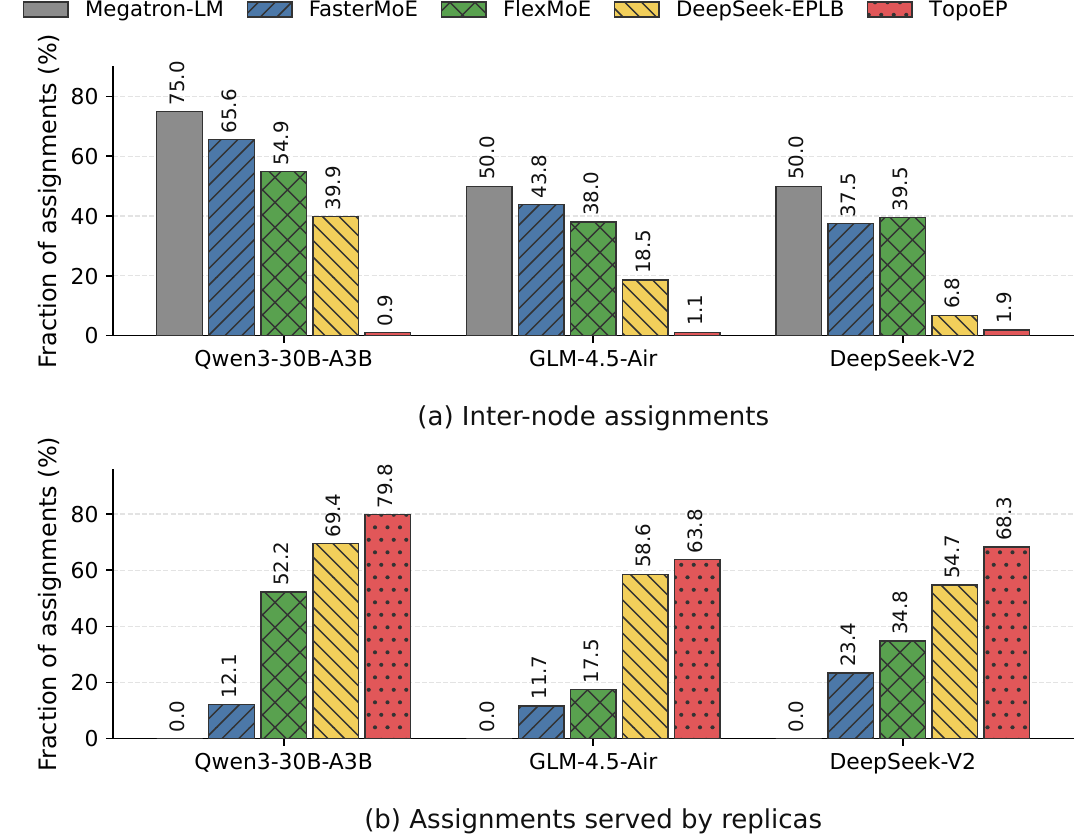}
    \caption{Physical execution of token--expert assignments.
    (a) Fraction whose source and execution ranks
    are on different nodes. (b) Fraction executed by replicas.}
    \label{fig:topology-aware-routing}
\end{figure}

Figure~\ref{fig:topology-aware-routing}(a)--(b) shows that
\textit{TopoEP} reduces the inter-node assignment fraction
from 50.0\%--75.0\% with Megatron-LM to 0.91\%--1.88\%,
while replicas execute 63.8\%--79.8\% of assignments.
Across all models, it achieves the lowest inter-node fraction and the highest
replica-served fraction. Together, these
results indicate that \textit{TopoEP} redistributes
most assignments to replicas while keeping their execution
within the source nodes.

The comparison with DeepSeek-EPLB shows that high replica
usage alone does not ensure communication locality.
On GLM-4.5-Air, DeepSeek-EPLB and \textit{TopoEP} serve
58.6\% and 63.8\% of assignments through replicas, respectively,
yet their inter-node fractions are 18.5\% and 1.1\%.
This contrast highlights the importance of coordinating
replica placement with token allocation, consistent with
\textit{TopoEP}'s two-stage strategy of placing expert
instances near token sources and refining rank loads
within each node.

\section{Related Work} \label{sec: work}
\paragraph{Host-side load balancing.}
System-level EPLB designs commonly rely on host-side planning.
FasterMoE~\cite{he2022fastermoe} replicates overloaded experts via shadow experts, FlexMoE~\cite{nie2023flexmoe} dynamically expands, shrinks, and migrates expert replicas, and DeepSeek-EPLB~\cite{deepseekai2025eplb} places redundant experts using a heuristic based on estimated historical loads.
These approaches can improve expert utilization, but host-side planning and reconfiguration add CPU--GPU coordination and state-management overhead,
making fine-grained adaptation hard to run on the critical path of each microbatch.

\paragraph{GPU-native load balancing.}
More recent systems move load-balancing decisions onto the GPU.
UltraEP~\cite{wei2026ultraep} and MoonEP~\cite{moonep2026} use lightweight planners for per-layer, per-microbatch balancing and avoid the host--GPU round trip.
However, both are designed primarily for a single scale-up domain. Their planners neither model the heterogeneous communication costs of NVLink and RDMA nor optimize cross-domain replica placement, so they are not directly applicable to the inter-node EP setting considered here.
\textit{TopoEP} targets this setting with a planner that models the hierarchical scale-up and scale-out topology when deciding replica placement and token rerouting.

\section{Conclusion}\label{sec:conclusion}
This paper presents \textit{TopoEP}, a GPU-native, topology-aware
load-balancing system for large-scale MoE training. It uses a deterministic
two-stage GPU solver to balance rank loads while accounting for the
heterogeneous communication costs of scale-up and scale-out fabrics, and
applies each resulting plan without data-dependent host synchronization. When
integrated with Megatron-LM on a 32-GPU NVIDIA H800 cluster, it improves
end-to-end training throughput by 6.2\%--11.4\% across three representative
MoE models without observable degradation in training-loss convergence.


\bibliographystyle{plain}
\bibliography{refs/intro, refs/background, refs/eval}

@inproceedings{li2023lina,
  author    = {Li, Jiamin and Jiang, Yimin and Zhu, Yibo and Wang, Cong and Xu, Hong},
  title     = {Accelerating Distributed {MoE} Training and Inference with {Lina}},
  booktitle = {2023 USENIX Annual Technical Conference (USENIX ATC 23)},
  year      = {2023},
  pages     = {945--959},
  publisher = {USENIX Association},
  address   = {Boston, MA},
  month     = jul,
  isbn      = {978-1-939133-35-9},
  url       = {https://www.usenix.org/conference/atc23/presentation/li-jiamin}
}

@InProceedings{pmlr-v235-kim24w,
  title = 	 {Scaling Beyond the {GPU} Memory Limit for Large Mixture-of-Experts Model Training},
  author =       {Kim, Yechan and Lim, Hwijoon and Han, Dongsu},
  booktitle = 	 {Proceedings of the 41st International Conference on Machine Learning},
  pages = 	 {24342--24353},
  year = 	 {2024},
  editor = 	 {Salakhutdinov, Ruslan and Kolter, Zico and Heller, Katherine and Weller, Adrian and Oliver, Nuria and Scarlett, Jonathan and Berkenkamp, Felix},
  volume = 	 {235},
  series = 	 {Proceedings of Machine Learning Research},
  month = 	 {21--27 Jul},
  publisher =    {PMLR},
  url = 	 {https://proceedings.mlr.press/v235/kim24w.html}
}

@inproceedings{vaswani2017attention,
  title     = {Attention Is All You Need},
  author    = {Vaswani, Ashish and Shazeer, Noam and Parmar, Niki and
               Uszkoreit, Jakob and Jones, Llion and Gomez, Aidan N. and
               Kaiser, {\L}ukasz and Polosukhin, Illia},
  booktitle = {Advances in Neural Information Processing Systems},
  volume    = {30},
  pages     = {5998--6008},
  publisher = {Curran Associates, Inc.},
  year      = {2017},
  url       = {https://papers.nips.cc/paper/7181-attention-is-all-you-need}
}

@inproceedings{10.1145/3503221.3508417,
author = {Ma, Zixuan and He, Jiaao and Qiu, Jiezhong and Cao, Huanqi and Wang, Yuanwei and Sun, Zhenbo and Zheng, Liyan and Wang, Haojie and Tang, Shizhi and Zheng, Tianyu and Lin, Junyang and Feng, Guanyu and Huang, Zeqiang and Gao, Jie and Zeng, Aohan and Zhang, Jianwei and Zhong, Runxin and Shi, Tianhui and Liu, Sha and Zheng, Weimin and Tang, Jie and Yang, Hongxia and Liu, Xin and Zhai, Jidong and Chen, Wenguang},
title = {BaGuaLu: targeting brain scale pretrained models with over 37 million cores},
year = {2022},
isbn = {9781450392044},
publisher = {Association for Computing Machinery},
address = {New York, NY, USA},
url = {https://doi.org/10.1145/3503221.3508417},
doi = {10.1145/3503221.3508417},
booktitle = {Proceedings of the 27th ACM SIGPLAN Symposium on Principles and Practice of Parallel Programming},
pages = {192–204},
numpages = {13},
location = {Seoul, Republic of Korea},
series = {PPoPP '22}
}

@InProceedings{pmlr-v162-liu22g,
  title = 	 {Gating Dropout: Communication-efficient Regularization for Sparsely Activated Transformers},
  author =       {Liu, Rui and Kim, Young Jin and Muzio, Alexandre and Hassan, Hany},
  booktitle = 	 {Proceedings of the 39th International Conference on Machine Learning},
  pages = 	 {13782--13792},
  year = 	 {2022},
  editor = 	 {Chaudhuri, Kamalika and Jegelka, Stefanie and Song, Le and Szepesvari, Csaba and Niu, Gang and Sabato, Sivan},
  volume = 	 {162},
  series = 	 {Proceedings of Machine Learning Research},
  month = 	 {17--23 Jul},
  publisher =    {PMLR},
  url = 	 {https://proceedings.mlr.press/v162/liu22g.html}
}

@article{zhou2022mixtureofexpertsexpertchoicerouting,
  title={Mixture-of-experts with expert choice routing},
  author={Zhou, Yanqi and Lei, Tao and Liu, Hanxiao and Du, Nan and Huang, Yanping and Zhao, Vincent and Dai, Andrew M and Le, Quoc V and Laudon, James and others},
  journal={Advances in Neural Information Processing Systems},
  volume={35},
  pages={7103--7114},
  year={2022}
}

@inproceedings{ICLR2026_6ed5bf44,
 author = {Guo, Wentao and Mishra, Mayank and Cheng, Xinle and Stoica, Ion and Dao, Tri},
 booktitle = {International Conference on Learning Representations},
 editor = {C. Vondrick and B. Hariharan and C. Raffel and L. Pinto and D. Yang and A. Faust},
 pages = {67814--67839},
 title = {SonicMoE: Accelerating MoE with IO and Tile-aware Optimizations},
 url = {https://proceedings.iclr.cc/paper_files/paper/2026/file/6ed5bf446f59e2c6646d23058c86424b-Paper-Conference.pdf},
 volume = {2026},
 year = {2026}
}

@inproceedings{MLSYS2025_e27ea0cd,
 author = {Zhang, Shulai and Zheng, Ningxin and Lin, Haibin and Jiang, Ziheng and Bao, Wenlei and Jiang, Chengquan and Hou, Qi and Cui, Weihao and Zheng, Size and Chang, Li-Wen and Chen, Quan and Liu, Xin},
 booktitle = {Proceedings of Machine Learning and Systems},
 editor = {M. Zaharia and G. Joshi and Y. Lin},
 pages = {},
 publisher = {MLSys},
 title = {COMET: Fine-grained Computation-communication Overlapping for Mixture-of-Experts},
 url = {https://proceedings.mlsys.org/paper_files/paper/2025/file/e27ea0cd50b798ff8942caf9203f0992-Paper-Conference.pdf},
 volume = {7},
 year = {2025}
}

@misc{deepep2025,
      title={DeepEP: an efficient expert-parallel communication library},
      author={Chenggang Zhao and Shangyan Zhou and Liyue Zhang and Chengqi Deng and Zhean Xu and Yuxuan Liu and Kuai Yu and Jiashi Li and Liang Zhao},
      year={2025},
      publisher = {GitHub},
      howpublished = {\url{https://github.com/deepseek-ai/DeepEP}},
}

@misc{shoeybi2020megatronlmtrainingmultibillionparameter,
      title={Megatron-LM: Training Multi-Billion Parameter Language Models Using Model Parallelism}, 
      author={Mohammad Shoeybi and Mostofa Patwary and Raul Puri and Patrick LeGresley and Jared Casper and Bryan Catanzaro},
      year={2020},
      eprint={1909.08053},
      archivePrefix={arXiv},
      primaryClass={cs.CL},
      url={https://arxiv.org/abs/1909.08053}, 
}

@misc{qwen3technicalreport,
      title={Qwen3 Technical Report}, 
      author={Qwen Team},
      year={2025},
      eprint={2505.09388},
      archivePrefix={arXiv},
      primaryClass={cs.CL},
      url={https://arxiv.org/abs/2505.09388}, 
}

@misc{deepseekv2,
      title={DeepSeek-V2: A Strong, Economical, and Efficient Mixture-of-Experts Language Model}, 
      author={DeepSeek-AI},
      year={2024},
      eprint={2405.04434},
      archivePrefix={arXiv},
      primaryClass={cs.CL}
}

@misc{zeng2025glm45,
  title         = {{GLM-4.5}: Agentic, Reasoning, and Coding ({ARC})
                   Foundation Models},
  author        = {Zeng, Aohan and Lv, Xin and Zheng, Qinkai
                   and Hou, Zhenyu and Chen, Bin and others},
  year          = {2025},
  eprint        = {2508.06471},
  archivePrefix = {arXiv},
  primaryClass  = {cs.CL},
  doi           = {10.48550/arXiv.2508.06471},
  url           = {https://arxiv.org/abs/2508.06471}
}

@inproceedings{
  penedo2024the,
  title={The FineWeb Datasets: Decanting the Web for the Finest Text Data at Scale},
  author={Guilherme Penedo and Hynek Kydl{\'\i}{\v{c}}ek and Loubna Ben allal and Anton Lozhkov and Margaret Mitchell and Colin Raffel and Leandro Von Werra and Thomas Wolf},
  booktitle={The Thirty-eight Conference on Neural Information Processing Systems Datasets and Benchmarks Track},
  year={2024},
  url={https://openreview.net/forum?id=n6SCkn2QaG}
}

@misc{penedo2025fineweb2pipelinescale,
  title         = {{FineWeb2}: One Pipeline to Scale Them All---Adapting
                   Pre-Training Data Processing to Every Language},
  author        = {Penedo, Guilherme and Kydl{\'i}{\v{c}}ek, Hynek and
                   Sabol{\v{c}}ec, Vinko and Messmer, Bettina and
                   Foroutan, Negar and Kargaran, Amir Hossein and
                   Raffel, Colin and Jaggi, Martin and von Werra, Leandro and
                   Wolf, Thomas},
  year          = {2025},
  eprint        = {2506.20920},
  archivePrefix = {arXiv},
  primaryClass  = {cs.CL},
  url           = {https://arxiv.org/abs/2506.20920}
}

@article{li2023starcoder,
  title   = {{StarCoder}: May the Source Be with You!},
  author  = {Li, Raymond and Ben Allal, Loubna and Zi, Yangtian and
             Muennighoff, Niklas and Kocetkov, Denis and Mou, Chenghao and
             Marone, Marc and Akiki, Christopher and others},
  journal = {Transactions on Machine Learning Research},
  year    = {2023},
  eprint  = {2305.06161},
  archivePrefix = {arXiv},
  primaryClass  = {cs.CL},
  url     = {https://arxiv.org/abs/2305.06161}
}

@inproceedings{soldaini-etal-2024-dolma,
    title = "Dolma: an Open Corpus of Three Trillion Tokens for Language Model Pretraining Research",
    author = "Soldaini, Luca  and
      Kinney, Rodney  and
      Bhagia, Akshita  and
      Schwenk, Dustin  and
      Atkinson, David  and
      Authur, Russell  and
      Bogin, Ben  and
      Chandu, Khyathi  and
      Dumas, Jennifer  and
      Elazar, Yanai  and
      Hofmann, Valentin  and
      Jha, Ananya  and
      Kumar, Sachin  and
      Lucy, Li  and
      Lyu, Xinxi  and
      Lambert, Nathan  and
      Magnusson, Ian  and
      Morrison, Jacob  and
      Muennighoff, Niklas  and
      Naik, Aakanksha  and
      Nam, Crystal  and
      Peters, Matthew  and
      Ravichander, Abhilasha  and
      Richardson, Kyle  and
      Shen, Zejiang  and
      Strubell, Emma  and
      Subramani, Nishant  and
      Tafjord, Oyvind  and
      Walsh, Evan  and
      Zettlemoyer, Luke  and
      Smith, Noah  and
      Hajishirzi, Hannaneh  and
      Beltagy, Iz  and
      Groeneveld, Dirk  and
      Dodge, Jesse  and
      Lo, Kyle",
    editor = "Ku, Lun-Wei  and
      Martins, Andre  and
      Srikumar, Vivek",
    booktitle = "Proceedings of the 62nd Annual Meeting of the Association for Computational Linguistics (Volume 1: Long Papers)",
    month = aug,
    year = "2024",
    address = "Bangkok, Thailand",
    publisher = "Association for Computational Linguistics",
    url = "https://aclanthology.org/2024.acl-long.840/",
    doi = "10.18653/v1/2024.acl-long.840",
    pages = "15725--15788"
}

@techreport{peS2o,
    author = {Luca Soldaini and Kyle Lo},
    year = 2023,
    title = {{peS2o (Pretraining Efficiently on S2ORC) Dataset}},
    institution = {{Allen Institute for AI}},
    note = {ODC-By, \url{https://github.com/allenai/pes2o}}
}

@inproceedings{yu2025dapo,
  author = {Yu, Qiying and Zhang, Zheng and Zhu, Ruofei and Yuan, Yufeng and
            Zuo, Xiaochen and Yue, Yu and others},
  title = {{DAPO}: An Open-Source {LLM} Reinforcement Learning
           System at Scale},
  booktitle = {Advances in Neural Information Processing Systems},
  editor = {Belgrave, D. and Zhang, C. and Lin, H.
            and Pascanu, R. and Koniusz, P. and Ghassemi, M.
            and Chen, N.},
  volume = {38, Main Conference},
  pages = {113222--113244},
  publisher = {Curran Associates, Inc.},
  year = {2025},
  doi = {10.52202/085713-3775},
  url = {https://proceedings.neurips.cc/paper_files/paper/2025/file/a4277440d50f1f15d2cb4c14f7e0c0d2-Paper-Conference.pdf}
}

@inproceedings{shazeer2017outrageously,
  author    = {Noam Shazeer and Azalia Mirhoseini and Krzysztof Maziarz and Andy Davis and Quoc V. Le and Geoffrey E. Hinton and Jeff Dean},
  title     = {Outrageously Large Neural Networks: The Sparsely-Gated Mixture-of-Experts Layer},
  booktitle = {5th International Conference on Learning Representations (ICLR)},
  year      = {2017},
  url       = {https://arxiv.org/abs/1701.06538}
}

@inproceedings{lepikhin2021gshard,
  author    = {Dmitry Lepikhin and HyoukJoong Lee and Yuanzhong Xu and Dehao Chen and Orhan Firat and Yanping Huang and Maxim Krikun and Noam Shazeer and Zhifeng Chen},
  title     = {{GShard}: Scaling Giant Models with Conditional Computation and Automatic Sharding},
  booktitle = {9th International Conference on Learning Representations (ICLR)},
  year      = {2021},
  url       = {https://openreview.net/forum?id=qrwe7XHTmYb}
}

@article{fedus2022switch,
  author  = {William Fedus and Barret Zoph and Noam Shazeer},
  title   = {Switch Transformers: Scaling to Trillion Parameter Models with Simple and Efficient Sparsity},
  journal = {Journal of Machine Learning Research},
  year    = {2022},
  volume  = {23},
  number  = {120},
  pages   = {1--39},
  url     = {https://www.jmlr.org/papers/v23/21-0998.html}
}

@inproceedings{dai2024deepseekmoe,
  author    = {Damai Dai and Chengqi Deng and Chenggang Zhao and R. X. Xu and Huazuo Gao and Deli Chen and Jiashi Li and Wangding Zeng and Xingkai Yu and Y. Wu and Zhenda Xie and Y. K. Li and Panpan Huang and Fuli Luo and Chong Ruan and Zhifang Sui and Wenfeng Liang},
  title     = {{DeepSeekMoE}: Towards Ultimate Expert Specialization in Mixture-of-Experts Language Models},
  booktitle = {Proceedings of the 62nd Annual Meeting of the Association for Computational Linguistics (Volume 1: Long Papers)},
  year      = {2024},
  pages     = {1280--1297},
  publisher = {Association for Computational Linguistics},
  doi       = {10.18653/v1/2024.acl-long.70},
  url       = {https://aclanthology.org/2024.acl-long.70/}
}

@misc{deepseekai2024deepseekv3,
  author        = {{DeepSeek-AI}},
  title         = {{DeepSeek-V3} Technical Report},
  year          = {2024},
  eprint        = {2412.19437},
  archivePrefix = {arXiv},
  primaryClass  = {cs.CL},
  url           = {https://arxiv.org/abs/2412.19437}
}

@inproceedings{gale2023megablocks,
  author    = {Trevor Gale and Deepak Narayanan and Cliff Young and Matei Zaharia},
  title     = {{MegaBlocks}: Efficient Sparse Training with Mixture-of-Experts},
  booktitle = {Proceedings of Machine Learning and Systems},
  year      = {2023},
  volume    = {5},
  pages     = {288--304},
  publisher = {MLSys},
  url       = {https://proceedings.mlsys.org/paper_files/paper/2023/hash/5a54f79333768effe7e8927bcccffe40-Abstract-mlsys2023.html}
}

@inproceedings{rajbhandari2022deepspeedmoe,
  author    = {Samyam Rajbhandari and Conglong Li and Zhewei Yao and Minjia Zhang and Reza Yazdani Aminabadi and Ammar Ahmad Awan and Jeff Rasley and Yuxiong He},
  title     = {{DeepSpeed-MoE}: Advancing Mixture-of-Experts Inference and Training to Power Next-Generation {AI} Scale},
  booktitle = {Proceedings of the 39th International Conference on Machine Learning},
  year      = {2022},
  volume    = {162},
  series    = {Proceedings of Machine Learning Research},
  pages     = {18332--18346},
  publisher = {PMLR},
  url       = {https://proceedings.mlr.press/v162/rajbhandari22a.html}
}

@inproceedings{hwang2023tutel,
  author    = {Changho Hwang and Wei Cui and Yifan Xiong and Ziyue Yang and Ze Liu and Han Hu and Zilong Wang and Rafael Salas and Jithin Jose and Prabhat Ram and HoYuen Chau and Peng Cheng and Fan Yang and Mao Yang and Yongqiang Xiong},
  title     = {{Tutel}: Adaptive Mixture-of-Experts at Scale},
  booktitle = {Proceedings of Machine Learning and Systems},
  year      = {2023},
  volume    = {5},
  pages     = {269--287},
  publisher = {MLSys},
  url       = {https://proceedings.mlsys.org/paper_files/paper/2023/hash/5616d34cf8ff73942cfd5aa922842556-Abstract-mlsys2023.html}
}

@article{wang2024auxiliary,
  author  = {Lean Wang and Huazuo Gao and Chenggang Zhao and Xu Sun and Damai Dai},
  title   = {Auxiliary-Loss-Free Load Balancing Strategy for Mixture-of-Experts},
  journal = {arXiv preprint arXiv:2408.15664},
  year    = {2024},
  url     = {https://arxiv.org/abs/2408.15664}
}

@article{nie2023flexmoe,
  author  = {Xiaonan Nie and Xupeng Miao and Zilong Wang and Zichao Yang and Jilong Xue and Lingxiao Ma and Gang Cao and Bin Cui},
  title   = {{FlexMoE}: Scaling Large-Scale Sparse Pre-Trained Model Training via Dynamic Device Placement},
  journal = {Proceedings of the ACM on Management of Data},
  year    = {2023},
  volume  = {1},
  number  = {1},
  pages   = {1--19},
  doi     = {10.1145/3588964},
  url     = {https://doi.org/10.1145/3588964}
}

@inproceedings{he2022fastermoe,
  author    = {Jiaao He and Jidong Zhai and Tiago Antunes and Haojie Wang and Fuwen Luo and Shangfeng Shi and Qin Li},
  title     = {{FasterMoE}: Modeling and Optimizing Training of Large-Scale Dynamic Pre-Trained Models},
  booktitle = {Proceedings of the 27th ACM SIGPLAN Symposium on Principles and Practice of Parallel Programming},
  year      = {2022},
  pages     = {120--134},
  publisher = {ACM},
  doi       = {10.1145/3503221.3508418},
  url       = {https://doi.org/10.1145/3503221.3508418}
}

@inproceedings{liu2023janus,
  author    = {Juncai Liu and Jessie Hui Wang and Yimin Jiang},
  title     = {{Janus}: A Unified Distributed Training Framework for Sparse Mixture-of-Experts Models},
  booktitle = {Proceedings of the ACM SIGCOMM 2023 Conference},
  year      = {2023},
  pages     = {486--498},
  publisher = {ACM},
  doi       = {10.1145/3603269.3604869},
  url       = {https://doi.org/10.1145/3603269.3604869}
}

@misc{deepseekai2025eplb,
  author       = {{DeepSeek-AI}},
  title        = {{EPLB}: Expert Parallelism Load Balancer},
  year         = {2025},
  publisher    = {GitHub},
  howpublished = {\url{https://github.com/deepseek-ai/EPLB}},
}

@article{hamidouche2025gpu,
  title={GPU-initiated networking for NCCL},
  author={Hamidouche, Khaled and Bachan, John and Markthub, Pak and Gootzen, Peter-Jan and Agostini, Elena and Jeaugey, Sylvain and Shafi, Aamir and Theodorakis, Georgios and Venkata, Manjunath Gorentla},
  journal={arXiv preprint arXiv:2511.15076},
  year={2025}
}

@misc{nvidia_hopper_2022,
    author       = {NVIDIA},
    title        = {{NVIDIA Hopper Architecture In-Depth}},
    howpublished = {\url{https://developer.nvidia.com/blog/nvidia-hopper-architecture-in-depth/}},
    year         = {2022},
}

@inproceedings{
nguyen2026leastloaded,
title={Least-Loaded Expert Parallelism: Load Balancing An Imbalanced Mixture-of-Experts},
author={Xuan-Phi Nguyen and Shrey Pandit and Austin Xu and Caiming Xiong and Shafiq Joty},
booktitle={Forty-third International Conference on Machine Learning},
year={2026},
url={https://openreview.net/forum?id=DFi6hHce5t}
}

@misc{wei2026ultraep,
  title         = {{UltraEP}: Unleash {MoE} Training and Inference on
                   Rack-Scale Nodes with Near-Optimal Load Balancing},
  author        = {Xinming Wei and Chao Jin and Tuo Dai and Yinmin Zhong
                   and Shan Yu and Chengxu Yang and Bingyang Wu and
                   Zili Zhang and Jing Mai and Qianchao Zhu and
                   Zhouyang Li and Yuliang Liu and Guojie Luo},
  year          = {2026},
  eprint        = {2606.04101},
  archivePrefix = {arXiv},
  primaryClass  = {cs.DC},
  doi           = {10.48550/arXiv.2606.04101}
}

@misc{moonep2026,
  author       = {Chen, Yutian and Li, Cong and Wang, Yucheng and Wei, Ming},
  title        = {{MoonEP}: A Perfectly Balanced Expert Parallelism Library via Dynamic Redundant Experts},
  year         = {2026},
  publisher    = {GitHub},
  howpublished = {\url{https://github.com/MoonshotAI/MoonEP}},
}

@inproceedings{skiadopoulos2026symi,
  title     = {{SYMI}: Efficient {Mixture-of-Experts} Training via Model and Optimizer State Decoupling},
  author    = {Skiadopoulos, Athinagoras and Zhao, Mark and Gandhi, Swapnil and Norrie, Thomas and Mukherjee, Shrijeet and Kozyrakis, Christos},
  booktitle = {23rd USENIX Symposium on Networked Systems Design and Implementation (NSDI 26)},
  pages     = {75--92},
  year      = {2026},
  month     = may,
  publisher = {USENIX Association},
  address   = {Renton, WA},
  isbn      = {978-1-939133-54-0},
  url       = {https://www.usenix.org/conference/nsdi26/presentation/skiadopoulos}
}

@misc{nie2022evomoeevolutionalmixtureofexpertstraining,
      title={EvoMoE: An Evolutional Mixture-of-Experts Training Framework via Dense-To-Sparse Gate}, 
      author={Xiaonan Nie and Xupeng Miao and Shijie Cao and Lingxiao Ma and Qibin Liu and Jilong Xue and Youshan Miao and Yi Liu and Zhi Yang and Bin Cui},
      year={2022},
      eprint={2112.14397},
      archivePrefix={arXiv},
      primaryClass={cs.LG},
      url={https://arxiv.org/abs/2112.14397}, 
}


\end{document}